\documentclass{aa}
 \pdfoutput=1

\usepackage{paralist,tikz}
\usetikzlibrary{shapes,arrows}

\usepackage[]{graphicx}
\usepackage{subfig}
\graphicspath{{./Figures/}}
\usepackage{booktabs}
\usepackage{amssymb}
\usepackage{epstopdf}
\usepackage{xcolor}
\usepackage{algorithmic}
\usepackage[lined,ruled,linesnumbered]{algorithm2e}

\usepackage[varg]{txfonts}
\usepackage{lscape}
\usepackage{geometry} 
\usepackage{siunitx}
\usepackage{hyperref}
\usepackage{natbib}
\usepackage[normalem]{ulem}
\bibpunct{(}{)}{;}{a}{}{,} 

\newcommand{\dm}{ }

\newcommand{\lya}{\ensuremath{\mathrm{Ly}\alpha}}
\newcommand{\lae}{\ensuremath{\mathrm{Ly}\alpha} emitters}
\newcommand{\ciii}[1]{[C\,\textsc{iii}]#1}
\newcommand{\oii}[1]{[O\,\textsc{ii}]#1}
\newcommand{\oiii}[1]{[O\,\textsc{iii}]#1}

\newcommand{\ha}{\ensuremath{\mathrm{H}\alpha}}

\newcommand{\ori}{\textsf{ORIGIN}}
\newcommand{\udft}{\textsf{udf-10}}

\newcommand{\beq}{\begin{equation}}
\newcommand{\eeq}{\end{equation}}
\newcommand{\beqna}{\begin{eqnarray}}
\newcommand{\eeqna}{\end{eqnarray}}

\newcommand{\bc}{{\bf{c}}}

\newcommand{\bx}{{\bf{x}}}

\newcommand{\bn}{{\bf{n}}}
\newcommand{\br}{{\bf{r}}}
\newcommand{\bs}{{\bf{s}}}
\newcommand{\bd}{{\bf{d}}}

\newcommand{\bSigma}{{\bf{\Sigma}}}

\newcommand{\bA}{{\bf{A}}}
\newcommand{\bF}{{\bf{F}}}

\newcommand{\bX}{{\bf{X}}}
\newcommand{\bC}{{\bf{C}}}
\newcommand{\bB}{{\bf{B}}}
\newcommand{\bD}{{\bf{D}}}
\newcommand{\bR}{{\bf{R}}}
\newcommand{\bN}{{\bf{N}}}
\newcommand{\bM}{{\bf{M}}}

\newcommand{\bS}{{\bf{S}}}

\newcommand{\bT}{{\bf{T}}}
\newcommand{\bW}{{\bf{W}}}

\newcommand{\btD}{{\widetilde{\bf{D}}}}

\newcommand{\balpha}{\boldsymbol{\alpha}}
\newcommand{\bepsilon}{\boldsymbol{\epsilon}}

\newcommand{\bsi}{{\bf{s}}_i}
\newcommand{\wbsi}{\widetilde{\bf{s}}_i}

\begin{document} 

 \title{ORIGIN: Blind detection of faint emission line galaxies in MUSE datacubes\thanks{\ori: \url{https://github.com/musevlt/origin}}}

 \author{
	   David Mary \inst{1}
	   \and Roland Bacon \inst{2}
	   \and Simon Conseil \inst{2}
       \and Laure Piqueras \inst{2}
       \and Antony Schutz \inst{1,2}
 }        

\institute{
 Universit\'e C\^ote d'Azur, Observatoire de la C\^ote d'Azur, CNRS, Laboratoire Lagrange, Bd de l'Observatoire, CS 34229, 06304, Nice cedex 4, France 
\and Univ Lyon, Univ Lyon1, Ens de Lyon, CNRS, Centre de Recherche Astrophysique de Lyon UMR5574, F-69230, Saint-Genis-Laval, France
   \\
    \email{david.mary@unice.fr}
}

\date{draft version 0.1}

\abstract
{One of the major science cases of the  MUSE (Multi Unit Spectroscopic Explorer) integral field spectrograph is the detection of Lyman-alpha emitters at high redshifts.  The on-going and planned deep fields observations  will allow for one large sample of these sources. An efficient tool to perform blind detection of faint emitters in MUSE datacubes is a prerequisite of such an endeavor. 
}
{Several line detection algorithms exist but their performance during the deepest MUSE exposures is hard to quantify, in particular with respect to their actual false detection rate, or purity.  \dm{The aim of this work is to design and validate} an algorithm that efficiently detects faint spatial-spectral emission signatures, while  allowing for a stable false detection rate over the data cube and providing in the same time an automated and reliable estimation of the purity. }
{The algorithm implements i) a nuisance removal part based on a continuum subtraction combining a discrete cosine transform and an iterative principal component analysis, ii) a detection part based on the local maxima of generalized likelihood ratio test statistics obtained for a set of spatial-spectral profiles of emission line emitters and iii)  a purity estimation part, where the proportion of true emission lines is estimated from the data itself:  the distribution of the local maxima in the ``noise only'' configuration is estimated from that of the local minima. }
{Results on simulated data cubes providing ground truth  show that the method reaches its aims in terms of purity and completeness.  When applied to the deep 30-hour exposure MUSE datacube in the Hubble Ultra Deep Field, the algorithms allows for the confirmed detection of 133 intermediate redshifts galaxies and 248 \lae, including 86 sources with no HST (Hubble Space Telescope) counterpart.}
{\dm{The algorithm fulfills its aims in terms of detection power and reliability. It is consequently}   implemented as a Python package whose  code and documentation are available  on  GitHub and readthedocs.}


\maketitle
%


\section{Introduction}
\label{sect:intro}
Spectroscopic observations of galaxies at high redshift have recently been revolutionized by the Multi Unit Spectroscopic Explorer (MUSE) instrument in operation at the VLT (Very Large Telescope) since 2014. MUSE is an adaptive optics assisted integral field spectrograph operating  in the visible \citep{Bacon+2010, Bacon+2014}. With its field of view of $1 \times 1$~arcmin$^2$ sampled at 0.2 arcsec and its simultaneous spectral range of 4800-9300~\AA\ at R$\sim$3000, MUSE produces large hyperspectral datacubes of 383 millions voxels, corresponding to about $ 320\times 320 \times 3680$ pixels along the spatial ($x,y$) and spectral ($z$) axes. Its unique capabilities of providing three-dimensional (3D) deep field observations have been demonstrated in the early observations of the Hubble Deep Field-South \citep{Bacon2015} and more recently in the Hubble Ultra Deep Field (HUDF, \cite{Bacon2017}) and the CANDELS - GOOD South area \citep{Urrutia2018}.

Thanks to its unrivalled capabilities, MUSE has been able to increase  the number of spectroscopic redshifts in these fields by an order of magnitude (see for example \citealt{Inami2017}). The most spectacular increase is at high redshift (z$>$3), where MUSE was able to detect a large number of \lae. In the deepest exposures (10+ hours), MUSE is even able to go beyond the limiting magnitude of the  deepest HST exposures. For example in the HUDF, which achieves a $5\sigma$ depth of 29.5 in the F775W filter, MUSE was able to detect \lae\ without an HST counterpart \citep{Bacon2017, Maseda2018}.
These observations have led to a breakthrough in our understanding of the high redshift Universe, which includes, for example, the discovery of \lya\ emission from the circumgalactic medium around individual galaxies \citep{Wisotzki2016, Leclercq2017, Wisotzki2018} or the role of the low mass galaxies and the evolution of the faint-end \lya\ luminosity function \citep{Drake2017}.

Building a large sample of low luminosity \lae\ at high redshift is an important objective for the near future with the upcoming deep fields observations currently executed or planned in the context of the MUSE \dm{guaranteed time observations (GTO)}. The availability of an efficient tool to perform blind detection of faint emitters in MUSE datacubes is a prerequisite of such an endeavor. 

Several tools have already been developed to perform blind searches of faint emitters in MUSE datacubes, such as \textsf{MUSELET}, a \textsf{SExtractor} based method available in \textsf{MPDAF} \citep{Piqueras2016}, \textsf{LSDCAT}, a matched filtering method \citep{LSDCAT} or \textsf{SELFI}, a Bayesian method \citep{2016A&A...588A.140M}.
These tools have been successfully used so far,  for instance  \textsf{LSDCAT} and \textsf{MUSELET} in the context of respectively the MUSE Wide Survey \citep{Urrutia2018} and the analysis of the lensing clusters (e.g., \cite{Lagattuta2019}). However, none of these methods currently {\dm{allow}} for a reliable estimate of the proportion of  false discoveries (or purity) actually present in the list of detected sources. As a consequence their actual performance on the deepest MUSE exposures, for which  no ground truth is available indeed, is consequently hard to quantify. 

Furthermore, from our experience in the blind search of emitters in the MUSE deep fields, we have learned that when tuned to be efficient for the faintest emitters, every detection method becomes overly  sensitive to any residuals left by the imperfect continuum subtraction of bright sources and by the data reduction pipeline (e.g., instrumental signatures or sky subtraction residuals). This leads to a global inflation  of the false detections, at levels
that are unpredictable and fluctuating in  the datacube. This effect requires either to limit the detection power by setting a threshold high enough to stay on the `safe side', or to consent spending a significant human-expert time to examine the long list of potential discoveries proposed by the algorithm.  

In this context, we have developed an automated method, called  \ori,  allowing for these methodological and computational challenges.  In this paper, we present in detail  the algorithm. \dm{ An overview is given in  Sec. \ref{sect:overview} and a step-by-step description in Sec.  \ref{sect:stepbystep}. The application of \ori \, to the deep 30-hours exposure MUSE datacube in the HUDF called \udft \; is presented in Sec. \ref{sect:udf}. Mathematical complements, implementation of the method into a publicly released software and parameters values used for the data cube \udft \, are given the Appendices}. Conclusions and possible improvements are listed in the last section.  
\section{Overview}
\label{sect:overview}
\begin{figure}
\centering
\tikzstyle{decision} = [diamond, draw, fill=blue!20, text width=5em, text badly centered, node distance=3cm, inner sep=4pt]
\tikzstyle{block} = [rectangle, draw, fill=blue!20, text width=7em, text centered, rounded corners, minimum height=4em]
\tikzstyle{line} = [draw, -latex']
\tikzstyle{cloud} = [draw, ellipse,fill=red!20, node distance=2cm, minimum height=1em]
   
\begin{tikzpicture}[node distance = 2cm, auto]
\node [cloud, text width=5em, text badly centered] (input) {Data cube \& variance  };
\node [block, below  of=input,text width=12em, node distance=2.5cm] (dct) {Nuisance removal : DCT \\ (Sec. \ref{sec:dct})};
\node [block, below  of=dct, node distance=3.1cm] (seg) {Segmentation \\ (Sec. \ref{sec:spatial})};
\node [block, right of=seg,  node distance=3.5cm] (pca) {Nuisance removal : PCA \\ (Sec. \ref{sec:PCA})};
\node [block, below of=pca, ,text width=9em, node distance=2.5cm] (glr) {Test statistics : GLR \\ (Sec. \ref{sec:GLR})};
\node [block, text width=10em, below left of=seg,node distance=3.5cm] (maxloc) {Test statistics : local extrema \\ (Sec. \ref{sec:locex})};
\node [block, below left of =glr, ,text width=10em, node distance=4.9cm] (purity) {Purity estimation and line detection \\ (Sec. \ref{sec:purity} and \ref{sec:pre})};
\node [block, below of=purity, node distance=2.5cm] (merging) {Line merging and sources' extraction \\ (Sec. \ref{sec:lines})};
\node [cloud, text width=5em, text badly centered, below  of=merging,node distance=2cm] (output) {Sources Catalog };
 
\path [line] (input) -- node {$\bf{S}$, $ \bf{\Sigma}$} (dct);
\path [line] (dct) -- node [near start] {$\bf{R}$, $\bf{\widehat{C}}$} (seg); 
\path [line] (seg) --  node { $\bf{L}$} (pca);
\path [line] (pca) --  node { $\bf{F}$} (glr);
\path [line] (glr) --  node { ${\bT}_{GLR}^+$, ${\bT}_{GLR}^-$} (maxloc);
\path [line] (maxloc) |- node [near start] { $\bf{M}^+$, $\bf{M}^-$} (purity);
\path [line] (purity) -- (merging);
\path [line] (merging) -- (output);
\path [line] (dct)  -| node [near end] {$\bf{R}$} (pca);
\path [line] (dct) -| node [near end] {$\bf{R}$} (maxloc); 
\end{tikzpicture}
\caption{\dm{Overview of the \ori \; algorithm. A session example is given in App. \ref{sect:implementation} and a list of the main parameters in App. \ref{annex:parameters}.}} \label{fig:M1}
\end{figure}
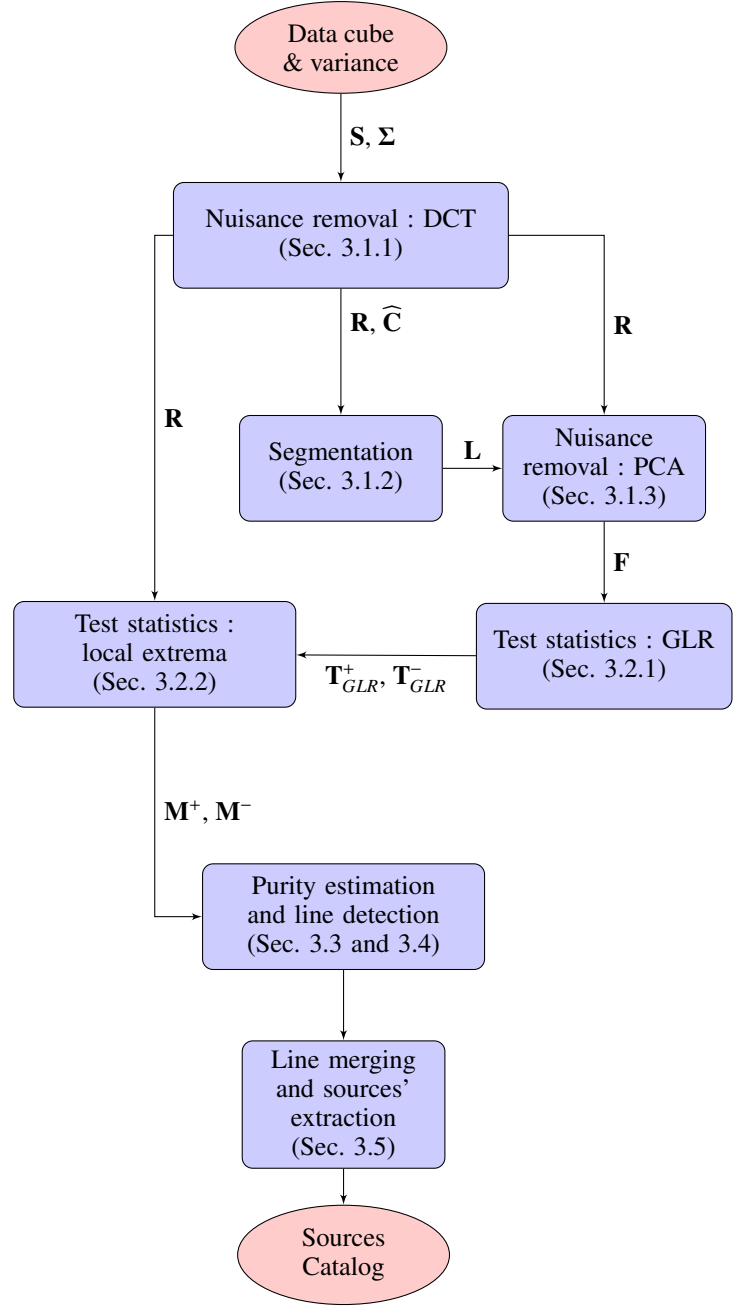
 
\subsection{\dm{Notations}}
\label{subsec:notations}
 In the following, we note  $N_x, N_y$ and $N_\lambda $  the two spatial and the spectral dimensions of the data cube. (For \udft\, this corresponds to $322\times 323\times 3681 \approx 383$ millions voxels.) Bold lowercase letters as $\bx$  denote vectors and bold capital letters as $\bX$  denote matrices. We also use the following notations: 
 \begin{itemize}
     \item $\bsi=[s_1,\hdots,s_{N_\lambda}]^\top$\quad is one spectrum  in the data cube, with $i\in\{1,\hdots, N_x\times N_y\}$,
     \item A collection of  spectra $\{{\bf{s}}_i\}$ from a data cube can be stored in a matrix whose columns are the spectra. For the whole data cube, this matrix is noted $\bS:=[\bs_1,\hdots, \bs_{N_x\times N_y}]$,
    \item $ \bSigma_i$ is the covariance matrix of  spectrum $\bsi$. It is provided by the data reduction pipeline.
      \item  Whitened (or standardized) vectors and matrices are denoted with a\; $\widetilde{ }$\;. For instance : $\wbsi:=\bSigma_i^{-\frac{1}{2}}\bsi$.
     \item  $ {\bf{0}}_{N_x,N_y,N_z}$ and  ${\bf{1}}_{N_x,N_y}$ represent
respectively  arrays of zeroes and ones of  size $N_x\times N_y \times N_z$ and
$N_x\times N_y $.
 \item The symbols $\otimes$ and $\oslash$ represent respectively  convolution and pointwise division.
  \item For a cube stored in a matrix $\bA$, notations $\bA(\cdot,\cdot,k)$ and $\bA(i,j,\cdot)$ represent respectively the image (`slice')
of the cube  in the $k^{\textrm{th}}$ wavelength band and the spectrum of the cube at spatial positions $(i,j)$.
     \end{itemize}
     For simplicity, we  often drop index $i$ when the described processing is applied sequentially to all spectra.
     
\subsection{\dm{Objectives}} 
The detection algorithm \ori\  is aimed at detecting point-like or weakly resolved emission line sources with emission lines   at unknown \dm{redshifts}. The algorithm was designed with a threefold objective : 1) Be as powerful as possible for detecting faint emission lines; 2) Be robust to local variations
of the data statistics, caused for instance by bright, extended sources, residual instrumental artifacts or different number of co-added exposures\dm{;} 3) Provide a list of detected sources that achieves a target purity, this
target purity being specified by the user.
These objectives led us to a detection approach in five main stages outlined below and described step by step in Sec. \ref{sect:stepbystep}.

\subsection{Overview}
\dm{  The flowchart in Fig. \ref{fig:M1} shows  an overview of the algorithm. We give here a brief summary of each step (purple boxes),  a detailed description of which can be found in the Sections indicated in these boxes. }
\subsubsection{Nuisance removal and  segmentation}
\label{sec:nuis_over}
Spatial and spectral components that are not
emission lines \dm{in the datacube $\bf{S}$} (e.g.,  bright sources, diffuse sources, instrumental artifacts) can drastically increase the false detection rate.
The first step is thus to detect and to remove such sources, called `nuisances' below.
Regions of the field that are free from such sources are called `background'. 
\dm{Nuisance features are removed 
by  combining a continuum removal (using a Discrete Cosine Transform, DCT) and an iterative Principal Component Analysis (PCA). The estimated continuum ($\bf{C}$) and the residual data cubes ($\bf{R}=\bf{S}-\bf{\widehat{C}}$)  from the DCT are used to perform a spatial segmentation. This provides for each spectrum a  label (nuisance or background. These labels are  stored in a matrix ($\bf{L}$)  and are used by the PCA, which acts differently on background and nuisance spectra.}
\subsubsection{Test statistics}
\dm{In the faint residual datacube ($\bf{F}$) the point is now to  capture  the signatures of line emissions}.   These signatures essentially take the form of `3D blobs' corresponding to the convolution of a spectral line profile by the spatial PSF of the instrument. Several matched filtering \citep{LSDCAT} or Generalized Likelihood Ratio (GLR) approaches \citep{Paris+2013,raja2013,ieee_sp_raja} have been devised for that purpose. Such approaches lead to filter  the  data cube with a library of possible signatures, built as spectral line profiles spread spatially by the PSF, and to normalize the result. 

\dm{The  filtering and normalization process considered here is also derived from a GLR approach  but it is robust to an unknown residual continuum  (see Sec. \ref{sec:test}). It produces two  data cubes (${\bT}_{GLR}^+$ and ${\bT}_{GLR}^-$), which correspond respectively to intermediate  tests statistics obtained when looking for emission and absorption lines. }  
When present in the data, an emission line emitter  tends to increase locally the values in ${\bT}_{GLR}^+$, with a local maximum close to the actual position (in $x$, $y$ and $z$ coordinates) of the line. 
These local maxima are used as final test statistics for line detection : each local maximum above a given threshold is identified as a line detection. 
It is important to note that for simplicity  ${\bT}_{GLR}^-$ is computed so that an absorption also appears as a local maximum in  ${\bT}_{GLR}^-$, 
so that   the final test statistics are  obtained as the local maxima  (called $\bf{M^+}$ and  $\bf{M^-}$)
of ${\bT}_{GLR}^+$ and ${\bT}_{GLR}^-$.
\subsubsection{Purity estimation}
Evaluating the purity of the detection results, that is, the fraction of true sources in the total list of detected sources, requires to estimate the number of local maxima that would be larger than the threshold `by chance', that is, in absence of any emission line.
 The complexity of the data prevents us from  looking for a  single, reliable analytical expressions of this distribution. However, the size of the data cube and the nature of our targets (emission lines) allows for the estimation of this distribution from the data - an approach advocated for instance by \cite{walter2016} and \cite{DBLP:journals/tsp/BacherMCM17}). When the data contains only noise and  under mild assumption on this noise, the statistics obtained when looking for emission lines \dm{
are the same as  those obtained when looking for absorption lines: the local maxima of $\bf{M^+}$ and $\bf{M-}$ should have the same distribution. Hence, the number of local maxima of $\bf{M^+}$ that should be above any given  threshold $\gamma$ under the null hypothesis  can be estimated from the  \dm{number of local maxima found above this threshold in $\bf{M^-}$}. This provides an estimate of the false discoveries that should be expected for any threshold, and hence of the purity}.
 In practice, this \dm{estimation} is done for \dm{a} grid of threshold values. The value of the threshold corresponding to the  purity desired by the user is identified and the  
 emission lines \dm{correspond to the local maxima of 
 $\bf{M^+}$} exceeding this threshold.  
\subsubsection{Detection of bright emission lines}
\dm{As explained in Sec. \ref{sec:nuis_over} the  iterative PCA  aims} at removing all features that do not correspond to faint emission lines. 
This means that 
bright emission lines --   \dm{the most easily detectable ones -- can be removed by the PCA step} and further be missed by the detection process. It is thus necessary to detect such lines before the PCA, in the DCT residual  ${\bf{R}}$. 
The procedure for {detecting \dm{these} lines and } controlling the purity {{of this detection step}} strictly mirrors what is done for faint lines, \dm{ with local maxima computed directly from \dm{(whitened versions of)} $\bf{R}$ and $-\bf{R}$ instead of ${\bT}_{GLR}^+$ and ${\bT}_{GLR}^-$}. 

\subsubsection{\dm{Line merging and sources' }extraction}
{\dm{The detected  bright and faint emission lines are finally merged into sources, for which several  informations (refined position, spectral profile and total flux) are computed and stored in the final source catalog (cf Sec. \ref{sec:lines}). 
}}

\section{Step-by-step method description}
\label{sect:stepbystep}

\subsection{Nuisance removal and spatial segmentation}
 The strategy of \ori \, is to track and suppress \dm{nuisance sources while preserving the targets, the line emitters. The  test statistics computed from  the residuals   still have to be robust against unknown fluctuating flux variations under the null hypothesis, but only moderately so if the nuisance removal is efficient (cf Sec.  \ref{sec:test}).}
 
 The removal of nuisance sources   is performed spectrum-wise in two steps explained below: DCT and iterative PCA. The first stage of DCT (a systematic and fast procedure) helps to remove quickly energetic and smooth fluctuations. In contrast, the version of the iterative PCA designed for this problem  can capture adaptively signatures that are much fainter and possibly very complex in shape, a but it is computationally heavy.  This \dm{combination} makes the overall nuisance removal process  efficient and tractable  in a reasonable time.
\subsubsection{Nuisance removal with DCT}
\label{sec:dct}
Each spectrum $\bs$ is modeled as
\beq 
 \bs = \bD \balpha + \bepsilon,\quad \textrm{ with}\quad \bepsilon\sim{\cal{N}}(\bf{0},\bSigma),
\eeq
where $\bD$ is a partial DCT matrix of size $N_\lambda \times N_{DCT}$, $\balpha$ is a $N_{DCT}\times 1$ vector of decomposition coefficients, $\bD \balpha$ is the unknown continuum and $\bepsilon$ is an additive Gaussian noise  with covariance matrix $\bSigma$. Maximum Likelihood estimation of the
continuum of spectrum $\bs$ leads to a
weighted least squares problem, for which the estimated
coefficients $\widehat{\balpha}$ are obtained by (cf sec. \ref{subsec:notations} for the\; $\widetilde{}$\; notation):
\beq
\widehat{\balpha} :=  \arg \displaystyle{\min_{\balpha}}\;\| \widetilde{\bs} -\btD \balpha \|^2 
 = (\btD^\top \btD)^{-1}\btD^\top \widetilde{\bs}.
\eeq
The estimated continuum ${\hat{\bc}}$ and residual  $\br$ are obtained by
\beq
\begin{cases}
\hat{\bc}= \bD \widehat{\balpha},\\
\br = \bs -  \hat{\bc}.
\end{cases}
\eeq
These spectra are collected in continuum and residual data cubes named $\widehat{\bC}$ and $\bR$ respectively.
The parameter $N_{DCT}$ controls the number of DCT basis vectors used in the continuum estimation. A value that is too small   lets large scale oscillations in the residual spectrum, while a value that is too large   tends to capture spectral features with small extension like  emission lines, which become then  more
difficult to detect in the residual. A satisfactory compromise was found here{\footnote{\dm{Note that the same trade-off must be accounted for when choosing the window length for a median filter, for instance.}}} with $N_{DCT}=10$. {\dm{This value leaves the lines almost intact: typically, the energy of the line in the DCT residual remains  close to 100\% until $N_{DCT}$ reaches several hundreds, depending on the line width. The continuum subtraction with $N_{DCT}=10$ is not perfect, but a large part of the work is done: \dm{for bright objects, 99 \% of the continuum's energy is typically contained in the subspace spanned by the first $10$ DCT modes and  decreases very slightly afterward.} The PCA does the rest. }}

Before describing the PCA we present  a segmentation step, which is required to implement the PCA.
\subsubsection{Spatial segmentation}
\label{sec:spatial}
The purpose of spatial segmentation 
is to locate regions where the data spectra contain nuisance sources and where they are free from them (in which case they are labeled as 'background'). This segmentation is necessary to further remove the nuisances. As mentioned before, such  \dm{spectra can have residuals from  continuum or bright emission lines}, or correspond to regions  exhibiting a particular statistical behavior, caused \dm{for instance} by  the presence of  instrumental artifacts or  residuals from sky subtraction .
The segmentation step relies both on the information extracted  in the continuum cube $\widehat{\bC}$ and on the  residual cube  $\widetilde{\bR}$
(which is whitened in order to account for the spectral dependence of the noise  variance). In $\widehat{\bC}$, an energetic spectrum $\widehat{\bc}$  indicates the presence of a continuum. In  $\widetilde{\bR}$,
a spectrum $\widetilde{\br}$ containing 
 residual signatures from bright sources or artifacts   tends to have higher energy than pure noise (background) pixels. 
For these reasons, we found that the following two 
tests statistics are both efficient and complementary to locate
nuisance sources:
\beq
\begin{cases}
t_1(\widehat{\bc}) := \log_{10} \|{\widehat{\bc}}\|^2,\\
t_2(\widetilde{\br}) := \frac{1}{N_\lambda}\|\widetilde{\br}\|^2.
\label{t1t2}
\end{cases}
\eeq
 For a spectrum under test $\bx$, the segmentation  tests  are both of the form 
\beq
t({\bf{x}}) {\displaystyle\mathop{\gtrless}_{\textrm{background}}^{\textrm{nuisance}}}
\gamma,
\label{test_t1t2}
\eeq
where $t$ is either $t_1$ or $t_2$ in \eqref{t1t2}, $\bx$ is either $\widehat{\bc}$ or $\widetilde{\br}$ and $\gamma$ is a threshold allowing for the tuning of
the sensitivity of the tests\footnote{For subsequent use (Algorithm 1), we    note  $\bf{t}_1(\bX)$ the vector collecting the test statistics of test $t_1$ applied to all spectra $\bx$ of cube $\bX$ (and similarly for ${\bf{t_2}}(\bX)$).}. In the data, the spectrum at   spatial coordinates $(i,j)$ in the field is considered containing nuisances if at least one
of the two tests applied to the corresponding spectra $\widehat{\bc}$ or $\widetilde{\br}$ \dm{leads to a result} above the test threshold.
 The $\log_{10}$ function in the definition of $t_1$ in \eqref{t1t2} is there to stabilize \dm{numerically}  the test statistic, as the squared norm of the estimated continuum may vary by several orders of magnitudes within a data cube. 
\begin{figure}
    \includegraphics[width=\columnwidth]{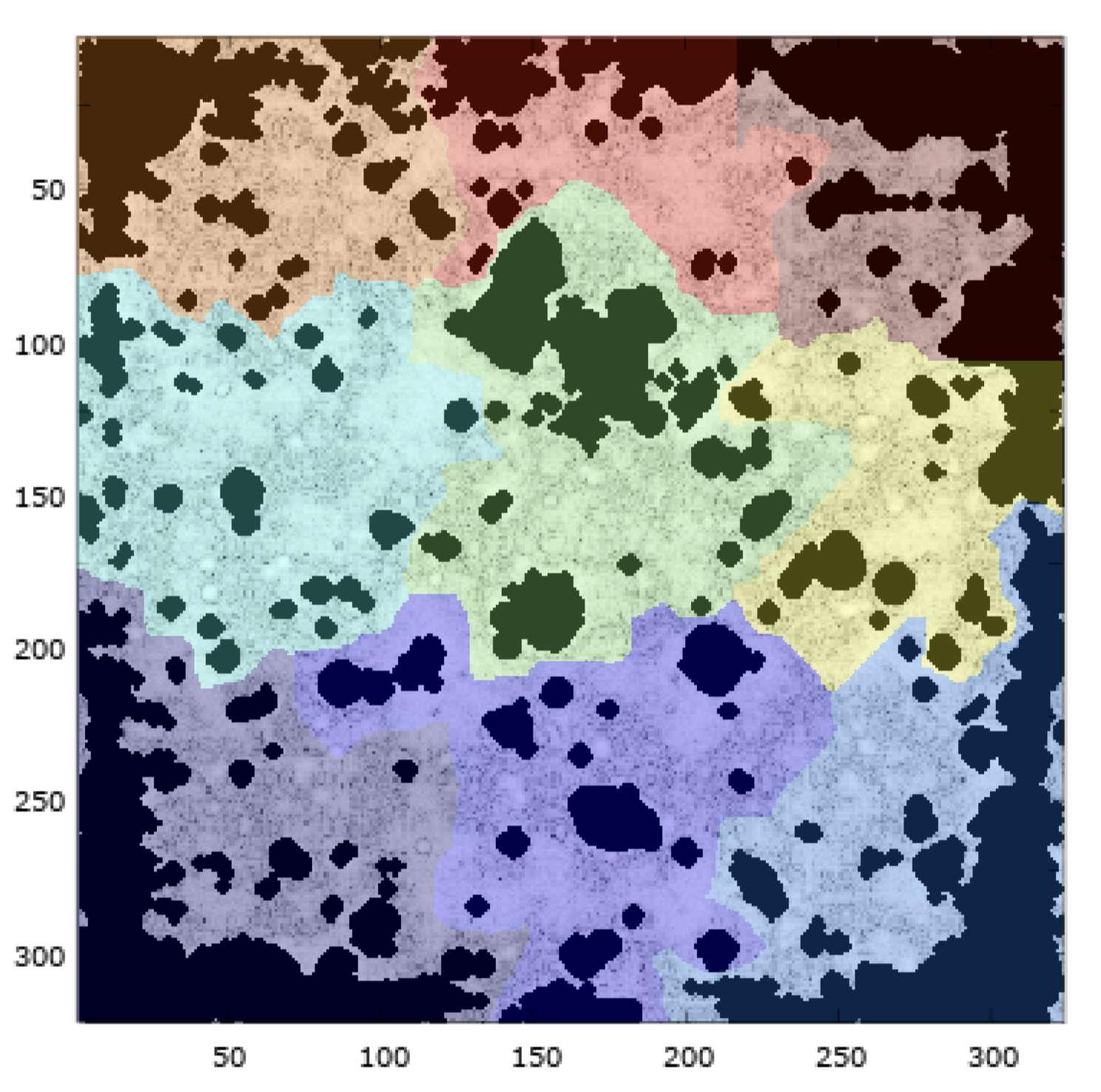}
    \caption{ Signal to Noise ratio (S/N) image (sum over the wavelength channels of the raw data cube divided by the standard deviation of the noise in each voxel) with, overlaid in black, the zones classified as nuisances by test $t_1$ and $t_2$. The large patches considered for the PCA removal are shown in color. }
    \label{fig:seg}
\end{figure}

\begin{figure}
     \includegraphics[width=\columnwidth]{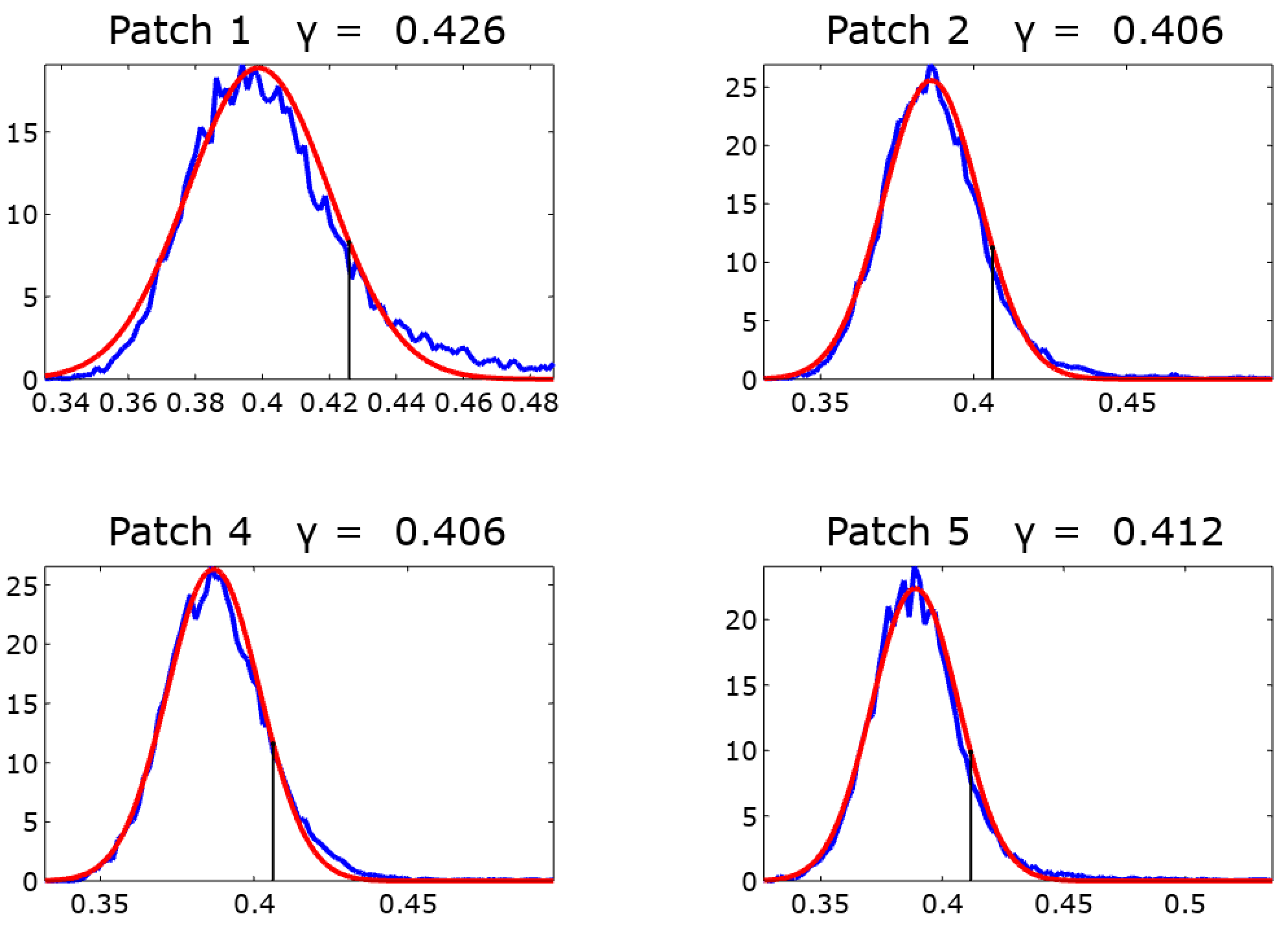}
    \caption{Distribution of the test statistics $t_2$ for four patches of $\widetilde{\bR}$ (blue), Gaussian approximation (red) and thresholds $\gamma(t_2)$ (in the titles and   black stems) above which all spectra of the patch are classified as nuisance for $P_{FA}^{PCA}= 0.1$.  The  plots for the five other patches are very similar and not shown.} 
    \label{fig:thresh}
\end{figure}

In  data like \udft, both distributions appear to be right-skewed, with a roughly Gaussian left tail  and a much heavier, possibly irregular right tail caused by nuisance sources. Hence, the distribution of the test statistics that would be obtained with pure noise (without nuisances) is estimated
by means of the left part of the empirical distribution, for which a Gaussian approximation provides a reasonable fit (see Fig.~\ref{fig:thresh}). This leads to an estimated  distribution of the test statistics under the noise only hypothesis which is Gaussian with estimated  mean $\widehat{\mu}(t)$ and estimated standard deviation $\widehat{\sigma}(t)$. For the purpose of  segmentation (or classification), the user chooses a  level of  error in the form of an error probability, called $P_{FA}^{seg}$ below. In order to tune the tests \eqref{test_t1t2} at this target error rate, one needs to find the threshold   value $\gamma$ such that 
$$ \textrm{Pr}\; (t>\gamma\;|\;\textrm{noise\; only})=P_{FA}^{seg},$$ 
The higher  the value of $P_{FA}^{seg}$, the lower the value of $\gamma$ and the larger the number of spectra wrongly classified as containing nuisances. If we denote by $\Phi$ the Cumulative Distribution Function (CDF) of a standard normal variable, the threshold $\gamma$ for $t_{1}$  is given by
\beq
\gamma(t) =\widehat{ {\mu}}(t)+\widehat{\sigma}(t) \cdot \Phi^{-1}(1-P_{FA}^{seg})  
\label{eq:gamma},
\eeq
where again $t$ is either $t_1$ or $t_2$.
 Two  segmented \dm{maps} are
obtained from thresholding the maps of the $N_x \times N_y$ test statistics  at the \dm{values } $\gamma(t_1)$ and   $\gamma(t_2)$ defined above. 
 \dm{The nuisances regions of each map are merged into a single merged segmentation map.}

Because MUSE data cubes are large, a PCA working on the full data cube would be sub-optimal. Indeed, the repeated computation of the eigenvectors of a matrix composed by the whole cube would be computationally prohibitive. Moreover, the aim of the PCA is to capture   spectral features corresponding to  nuisances. Such nuisances have   features that are locally similar, so a PCA working on  patches smaller than
 the whole cube is also more efficient to remove the nuisances. When building these  patches (\dm{whose size is } typically one tenth of the whole data cube for \udft), care must be taken that regions identified as nuisances in the previous step are not split
over two such patches. For \udft\, the segmentation algorithm starts with $N_Z=9$ rectangular patches. The nuisance zones intersecting several such patches are iteratively identified and attached to one patch, under the constraint that the final patches have surface in a target range. The minimal and maximal surfaces allowed for
patches of \udft\ are respectively $S_{min}=80^2$ and $S_{max}=120^2$ pixels$^2$.
 The results of the segmentation nuisance versus background for \udft\ after merging the two maps based on $t_1$ and $t_2$ is shown in Fig. \ref{fig:seg}.  The Figure also shows the segmentation \dm{result} into a number of $N_Z=9$ large patches.

\subsubsection{Nuisance removal with iterative PCA}
\label{sec:PCA}
The goal of this step is to iteratively locate and remove from  $\widetilde{\bR}$ residual nuisance sources, that is, any signal that is not the signature of a faint, spatially unresolved emission line. The algorithm cleans the cube  one patch at a time. 
In order to leave in the cleaned data cube a structure compatible with noise, the nuisances are constrained to leave outside \dm{a `noise subspace'}, which is \dm{defined as} the sample average of a small fraction ($F_b$) of the spectra flagged as `background' in the patch.\\ The Algorithm works as follows (see the pseudo-code  in Algorithm 1). At each iteration, the algorithm classifies spectra of the patch \dm{under consideration ($Z_i$)} as background or nuisance and stores them in matrices respectively called  $\bB$ and $\bN$. This classification is done  by means of the test  $t_2$ in \eqref{t1t2} - \eqref{test_t1t2}, applied to all the  spectra in the patch $Z_i$. For this test a value
 $P_{FA}^{PCA}$ is chosen by the user, and the corresponding test threshold is computed as in \eqref{eq:gamma} with $P_{FA}^{PCA}$ replacing $P_{FA}^{seg}$. In practice, for fields relatively empty as \udft,  values of  $P_{FA}^{PCA}$ and $P_{FA}^{seg}$ in the range $[0.1, 0.2]$ \dm{typically provide} a good trade-off in  cleaning nuisances without impacting Ly$\alpha$ emission lines. Fig. \ref{fig:thresh} shows in black the value of the
threshold for four patches $Z_i$ of the $\udft$ data cube.

From the fraction \dm{($F_b$ \%)} of the spectra that show the
lowest test statistics, a mean background $\overline{\bf{b}}$ is estimated and all the nuisance spectra in $\bN$ are orthogonalized with respect to this background.  This
results in a matrix of spectra $\overline{\bN}$, of which the first eigenvector is computed. The contribution of this vector to all spectra of  the patch is removed. If some of the resulting residual spectra are still classified as nuisances, the process is repeated until there is no more spectrum  classified as nuisance in the patch.

\begin{figure}
    \includegraphics[width=\columnwidth]{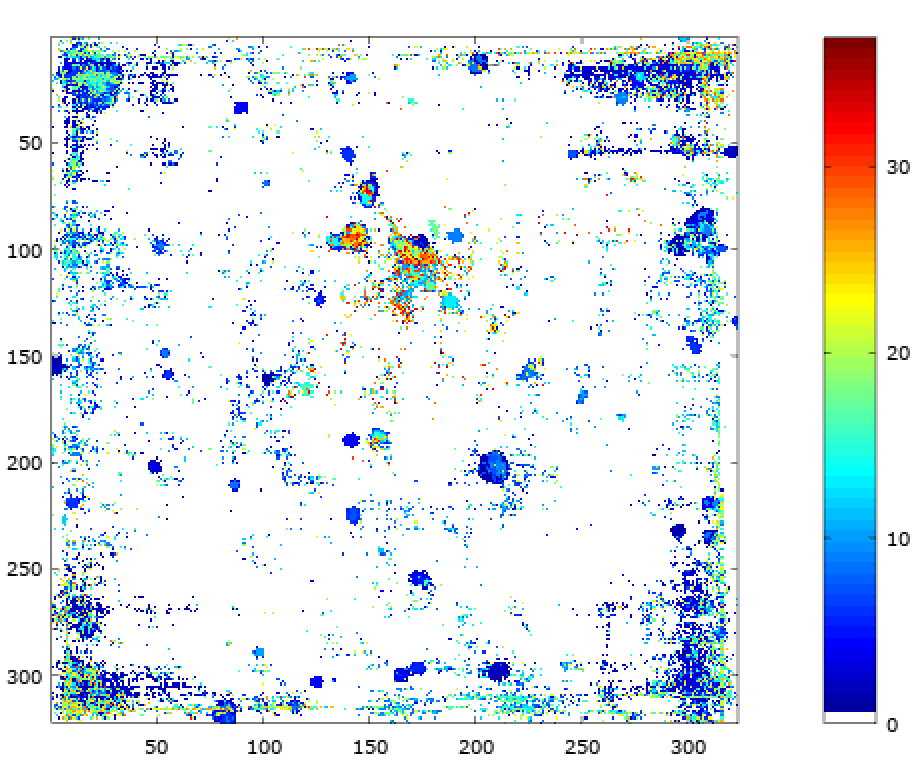}
    \caption{Iteration map of  greedy PCA for $\udft$ with $P_{FA}^{PCA}=0.1$. The map shows
    how many times each spectrum was selected in the nuisance matrix $\bN$ before all spectra were considered `cleaned' (that is, were classified as background by  test $t_2$).}
    \label{fig:iterationmap}
\end{figure}

Fig. \ref{fig:iterationmap} shows how many times each spectrum was classified as nuisance during the PCA cleaning stage. Comparing with the S/N image (gray and black zones in Fig.~\ref{fig:seg}), this iteration map clearly evidences regions where bright sources and artifacts \dm{ are present} (e.g., horizontal lines close to the borders in the top and bottom right corner).  Note also that the process is relatively fast with less than 40 iterations required in each patch to converge to a cleaned data cube.

\IncMargin{1em}
\begin{algorithm}
\SetKwInOut{Input}{Inputs}\SetKwInOut{Output}{Output}
\Input{$\widetilde{\bR}$,  \dm{$F_b$},  $P_{FA}^{PCA}$, empty data cube $\bF$.}
\BlankLine
\Output{`Cleaned' data cube $\bF$.}
\BlankLine
\For{$i\leftarrow 1$ \KwTo $N_z$}
{{ $\bF_i \leftarrow$\ set of spectra of $\bF$ in zone $i$}\\
{ $\bX \leftarrow$\ set of spectra of $\widetilde{R}$ in zone $i$}\\
  {Compute $\bB$ and $\bN$ using ${\bf{t}_2(\bX)}$ as in \eqref{t1t2}} \\
  \While{$\bN$ is not empty}
  {\label{forins}
   Estimate $\overline{\bf{b}}$ as the mean of the \dm{$F_b$ \% of the} spectra having the lowest test statistics in ${\bf{t}}_2$ \\
     $ \perp$ $ \bN$ with respect to  $\overline{\bf{b}}$    : $\overline{\bN}\leftarrow\bN- \frac{\overline{\bf{b}}\, \overline{\bf{b}}^\top}{\overline{\bf{b}}^\top \overline{\bf{b}}}\bf{N}$\\
  Compute the first eigenvector $\bf{v}$ of $\overline{\bN}$\\
   $ \perp$ $ \bX$ with respect to  ${\bf{v}}$  :
   ${\bX}\leftarrow\bX- {{\bf{v}}\, {\bf{v}}^\top}\bf{X}$\\
 Compute $\bB$ and $\bN$ using ${{\bf{t}}_2(\bX)}$ as in \eqref{t1t2}
} 
Spectra of $\bF_i \leftarrow\bX  $
}
\caption{Iterative greedy PCA algorithm}
\label{greedypca}
\end{algorithm}
\DecMargin{1em}


\subsection{Test statistics}
\label{sec:test}
\dm{\subsubsection{Generalized Likelihood Ratio}
\label{sec:GLR}}
For all  positions  $(x,y,z)$ in the PCA residual data cube $\bF$ (for `faint'), the algorithm considers sub-cubes ${\bf{f}}(x,y,z)$ of $\bF$ (called ${\bf{f}}$ for short below), centered on position
$(x,y,z)$ and having the size of the considered target signatures. For each such subcube we formulate a binary hypothesis test between 
\begin{itemize}
    \item 
 the null hypothesis: there is no  emission line centred at position $(x,y,z)$,
\item the alternative hypothesis: there is one  emission line $ \bd_i$ centred at position $(x,y,z)$, where $ \bd_i$ is a `3D' (spatial-spectral) signature among a set of $N_s$ possible line signatures.
\end{itemize}

\IncMargin{1em}
\begin{algorithm}
\SetKwInOut{Input}{Inputs}\SetKwInOut{Output}{Output}
\Input{${\bF}$,  $N_x,N_y,N_z$,$N_s$, dictionary of PSF $\{{\bf{p}}_i\}$, with $i=1,\hdots,N_z$, dictionary of spectral profiles $\{{\bf{d}}\}_i$ with $i=1,\hdots,N_p$. 
}
\BlankLine
\Output{ $\bT_{GLR}^+$,  $\bT_{GLR}^-$,  $\bM^+$ and $\bM^-$.}
\BlankLine
Define  $\bT_{GLR}^+ =  \bT_{GLR}^- = {\bf{0}}_{N_x,N_y,N_z}$\\
\For{$i\leftarrow 1$ \KwTo $N_z$}
{Subtract the mean $m_i$ of ${\bf{p}}_i$ to ${\bf{p}}_i$: ${\overline{\bf{p}}}_i \leftarrow {\bf{p}}_i - m_i {\bf{1}} $ \\
Compute: $\|  \overline{\bf{p}}_i \|^2 \leftarrow \overline{\bf{p}}_i ^\top \overline{\bf{p}}_i$ \\
$ \bW(\cdot,\cdot,i)\leftarrow \|  \overline{\bf{p}}_i \|^2{\bf{1}}_{N_x,N_y}$ \\
Convolve   band $i$ of $\bF$ with $\overline{\bf{p}}_i$ and store in
$\bT$:
$\bT(\cdot,\cdot,i)=\bF(\cdot,\cdot,i)\otimes  \overline{\bf{p}}_i$
}
\For{$i\leftarrow 1$ \KwTo $N_s$}
{
Subtract the mean $m_i$ of ${\bf{d}}_i$ to ${\bf{d}}_i$: ${\overline{\bf{d}}}_i \leftarrow {\bf{d}}_i - m_i {\bf{1}} $ \\
Compute the squared zero-mean profile,  ${\bf{d}}_i^{s}(n)$:  ${\bf{d}}_i^{s}(n) \leftarrow {\bf{d}}_i^ 2(n) $ for all entries ${\bf{d}}_i(n)$ of ${\bf{d}}_i$\\
\For{$j\leftarrow 1$ \KwTo $N_x$}
{\For{$k\leftarrow 1$ \KwTo $N_y$}
{
$\bT(i,j,\cdot)\leftarrow \bT(i,j,\cdot) \otimes  \overline{\bf{d}}_i$\\
$\bW(i,j,\cdot)\leftarrow \bW(i,j,\cdot) \otimes  \overline{\bf{d}}_i^s $
}
}
Normalize: $\bT\leftarrow \bT \oslash \bW $\\
\For{{\rm{all voxels}} $(i,j,k)$}
{\If{ $ {\bf{T}} > {\bf{T}}_{GLR}^+(i,j,k) $}{ $ {\bf{T}}_{GLR}^+(i,j,k) \leftarrow {\bf{T}}(i,j,k) $}
\If{$ {\bf{T}} < {\bf{T}}_{GLR}^-(i,j,k) $}{$ {\bf{T}}_{GLR}^-(i,j,k) \leftarrow {\bf{T}}(i,j,k) $}
}
}
${\bf{T}}_{GLR}^- \leftarrow - {\bf{T}}_{GLR}^- $\\
$\bM^+ \leftarrow$ local maxima $({\bf{T}}_{GLR}^+)$ \\
$\bM^- \leftarrow$ local maxima $({\bf{T}}_{GLR}^-)$ 
\caption{Computation of the test statistics }
\label{greedypca}
\end{algorithm}
\DecMargin{1em}

The statistical model retained to describe these two hypotheses is important, as it should capture as reliably as possible the statistical distribution of the data. This distribution results from a long chain of preprocessing, from  the data reduction pipeline to the PCA described in the previous step, and may thus
be very complex. On the other hand, a too sophisticated model may  lead to untractable processing because it  requires, in a GLR approach, maximum likelihood estimation of the  corresponding unknown parameters   for   380 millions positions in the datacube. We compared the performances of several statistical models and opted for the following, which allows for 
 a good compromise between computational efficiency, detection power and robustness to remaining faint artifacts:
\begin{equation}
\begin{cases}
{\mathcal{H}}_0: {\bf{f}} = a {\bf{1}} + \bn, \\
{\mathcal{H}}_1: {\bf{f}} = a {\bf{1}} + b \bd_i + \bn  ,  \textrm{\;with \;} i\in\{1,\hdots,N_s\} \textrm{\;unknown},
\label{mod}
\end{cases}
\end{equation}
where $\bn\sim \mathcal{N}(${\bf{0}}$, ${\bf{I}}$)$ is the noise assumed to be zero mean Gaussian with Identity covariance matrix under both hypotheses. The term $a \bf{1}$, with $a\in \mathbb{R}$ and $\bf{1}$ a vectorized cube of ones, models a possible residual and  unknown  nuisance flux, that is considered spatially and spectrally constant in subcube ${\bf{f}}$. The term $b \bd_i$, with $b>0$ and $i \in \{1,\hdots,N_s\}$, corresponds to one of the possible  emission signatures. Each signature is a spectral profile of a given width, spatially spread in each channel by the PSF in this channel.  
 The considered signatures  define the size of  ${\bf{f}}$. Spatially, they have the size of the PSF ($25\times 25$ for \udft). Spectrally, $N_s$ sizes  are considered. For the presented \udft\ analysis we used $N_s=3$ Gaussian profiles  with FWHM of 2, 6.7 and 12 spectral channels, covering respectively $5, 20$ and $ 31$ spectral channels in total.
 
 For binary hypothesis testing involving unknown parameters, the GLR approach (\cite{scharf94,kaydetection}) forms the test statistics by plugging in the likelihood ratio the maximum likelihood estimates of
 these parameters (namely, $a$ under ${\mathcal{H}}_0$ and $\{a,b,i\}$
 under ${\mathcal{H}}_1$). 
 This leads for each subcube {\bf{f}} to a test
 statistics  in the form of a matched filter (see Appendix \ref{deriv}):
\beq
T^+_{GLR}({\bf{f}}):=\max_i\left( 0,
\frac{{\bf{f}}^{\;\top} \overline{\bd} _i}{\|  \overline{\bd}_i \|} \right),
\label{tglr}
\eeq
where the  superscript $^+$ refers to positive (emission) lines and
 $\overline{\bd}_i$ denotes the spatial-spectral signature  ${\bd}_i$
 to which the mean has been subtracted. Eq. \eqref{tglr} can be efficiently
 computed for all positions $(x,y,z)$  using  Algorithm 2. The first main loop (rows 2 to 7) processes the cubes channel by channel. The second main loop (rows 8 to 26) processes the result of the first loop profile by profile, with an embedded loop processing  spectrum per spectrum (rows 11 to 16). Comparisons of the score obtained for each profile (rows 19 to 25) implement the max and min operators required for $T_{GLR}^+$ and  $T_{GLR}^-$ (cf Eqs. \eqref{tglr} above and \eqref{glrm} below).



\begin{figure}
    \includegraphics[width=\columnwidth]{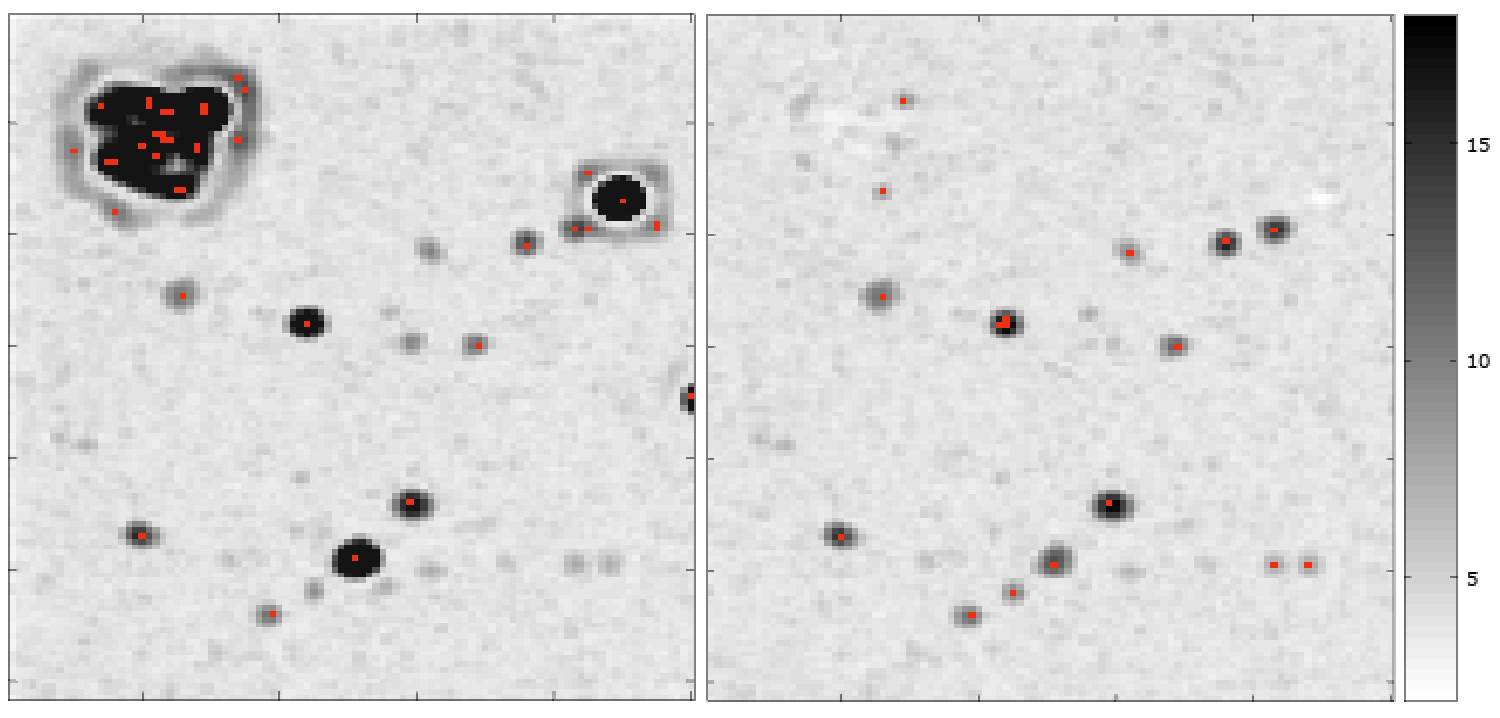}
    \caption{Comparison of two different  nuisance removal algorithms on the $100\times 100$  top left region of the \udft field.  Left : median filtering with a window of $144$ channels.  Right : DCT $+$ PCA. Red : local maxima $\bM^+$ larger than $v =$ $3 \sigma$ after the mean ($v=9.4$ for the median and $v=6.6$ for the DCT+PCA). The grayscale ranges from 0 (white) to 16 (black). }
    \label{comparcor}
\end{figure}

\dm{\subsubsection{Local extrema}
\label{sec:locex}}
The GLR test statistics result in a cube of test statistics values, ${\bT}_{GLR}^+$. 
When a line emission is present at a position $p_0 := (x_0,y_0,z_0)$, the values of  ${\bT}_{GLR}^+$ tend to 
increase \dm{statistically} in the vicinity of $p_0$ \dm{with} a local maximum at (or near) $p_0$.  For this reason, detection approaches based on local maxima are  often advocated,  possibly after a matched-filtering step, when \dm{`blobs'} of moderate extensions have to be detected in  random fields
(e.g., \cite{sobey92,bardeen86,2016A&A...589A..20V,2017A&A...604A.115V,peak-pvaluesSchwartzman358051}).
We opt for this approach and consider in the following the cube obtained by computing the three-dimensional local maxima of ${\bT}_{GLR}^+$, noted $\bM^+$ and defined by: 
\vspace{-2mm}
\beq\bM^+(x,y,z):=
\begin{cases}
 \bT_{GLR}^+(x,y,z)\;\;  \textrm{if}\;\;  \bT_{GLR}^+(x,y,z) > {\bf{v}}(i),\forall i=1\hdots,26,\\
0\quad \quad \quad \quad \quad \textrm{otherwise},
\label{M+}
\end{cases}
\eeq
where $\bf{v}$ collects the 26 voxels connected by faces, edges or corners touch to voxel $\bT^+_{GLR}(x,y,z)$. 

 An example of the resulting
 cube of test statistics is represented in Fig. \ref{comparcor}, right panel. This panel shows in gray scale the maximum over the spectral index of  $\bT^+_{GLR}$ obtained for a region of \udft. The corresponding local maxima (nonzero values of $\bM^+$) are shown in red. In this panel, 
  $\bT^+_{GLR}$ and $\bM^+$
 reveal very clearly regions  where spatially unresolved emission lines are likely to be present (darker \dm{blobs} of the \dm{size of the PSF}, and red points). In comparison, the left panel shows
 the GLR cube obtained when applying Algorithm 2  after a preprocessing based solely on a median filtering (instead of the DCT+PCA processing described above).
Clearly, a less efficient nuisance removal leads, in the vicinity of bright sources for instance, to wild and undesired variations of the test statistics.
 Such a loss of efficiency in the preprocessing stage would have to be compensated for by a possibly time consuming \dm{postprocessing} stage.

In order to evaluate how significant are the claimed detections, an important question is to know the distribution of the local maxima under ${\mathcal{H}}_0$. This problem has been studied  in applications like  the heights of waves in stationary sea state \citep{sobey92},  the fluctuations of the cosmic background \citep{bardeen86}, source detection in radiointerferometric maps \citep{2016A&A...589A..20V,2017A&A...604A.115V}  or brain activity regions in neuroimaging \citep{peak-pvaluesSchwartzman358051}, the random fields in which local maxima are considered being two-dimensional for the former three applications and three-dimensional for the latter.

In all these works, the distribution of the local maxima are derived from theoretical results regarding Gaussian random fields, see
in particular the works of \cite{cheng2018} for  recent results and a review.  We opted here for a different approach however, because the random field $\bT_{GLR}^+(x,y,z)$  is not guaranteed to be Gaussian nor smooth (owing for instance to the maximum over the spectral profiles computed in \eqref{tglr}). Besides, it is not isotropic and
 the correlation structure is likely to be not stationary in the spatial dimension owing to the varying number of exposures and to telluric spectral lines. While it might be possible  to derive an accurate statistical characterization of the local maxima of MUSE data by combining results of \cite{peak-pvaluesSchwartzman358051} for 3D, non isotropic  non-stationary and possibly non Gaussian random fields, this approach would deserve a far more involved study. We present and validate here 
a relatively more simple and empirical  approach. This approach combines inference based on local maxima with an estimation of the  distribution under ${\mathcal{H}_0}$ directly from the data itself, as considered for instance in  \cite{walter2016} and \cite{DBLP:journals/tsp/BacherMCM17} and further motivated here by the large dimension of the  data cubes. For this, our approach is in two steps.
\begin{itemize}
\item Step 1 : Compute GLR scores for {absorption} lines under ${\mathcal{H}_0}$.\\ 
Consider that we wish to detect absorption (rather than emission) lines. The model is the same as \eqref{mod} but with this time $b<0$. The GLR leads to the tests
statistics
\beq
T_{GLR}^-:=-\min_i\left(0,\frac{-{\bf{f}}^{\;\top} \overline{\bd} _i}{\|  \overline{\bd}_i \|}\right).
\label{glrm}
\eeq
As shown in Appendix \ref{app1},  under ${\mathcal{H}}_0$,  $T_{GLR}^+$ and $T_{GLR}^-$ share the same distribution.

\item{Step 2 : Statistics of the local maxima}. Since the distribution of $T^+_{GLR}$ and $T^-_{GLR}$ are the same, 
a training set of test statistics for the local maxima $\bM^+$ can be obtained by the local maxima of $T^-_{GLR}$, noted $\bM^-$, which is defined similarly as $\bM^+$ in \eqref{M+},
with $\bT^-$ replacing $\bT^+$. In practice, the background region in which
the statistics are estimated is obtained by the merged  segmentation \dm{maps}
obtained from tests $t_1$ and $t_2$ in \eqref{test_t1t2}. 
\end{itemize}

\subsection{Purity estimation}
\label{sec:purity}
The purity (or fidelity) is defined as the proportion of true discoveries (called $S$ below) with respect to the total
number of discoveries ($R$). With these notations the number $V:=R-S$ is the number of false discoveries and the purity is
\beq
P:=\frac{S}{R}=\frac{R-V}{R}=1-\frac{V}{R} ,
\label{pur}
\eeq
where $\frac{V}{R}$ is sometimes called False Discovery Proportion (FDP). We note that when a large number of comparisons is made in  multiple testing procedures (as this is the case here with millions of local maxima), 
controlling the false alarm (or familywise error) rate at low levels leads to drastically increase the test's threshold, which results in a substantial loss of power. In such situations it can be more efficient to control instead 
 the False Discovery Rate (FDR, \cite{bh1995}), defined as
\beq
\textrm{FDR}:=\textrm{E}\; (\textrm{FDP})= 1- \textrm{E} ({P}),
\label{fdr}
\eeq
where E$(\cdot)$ denotes expectation and by convention FDR$=0$ if $R=0$. Definition \eqref{fdr} shows that procedures aimed at controlling  purity and  FDR are indeed connected.

Coming back to our line detection problem, we wish to find a procedure making it possible 
to decide which local maxima
of $\bM^+$ should be selected so that the purity of the resulting selection is guaranteed at a level prescribed by the user. This amounts to find the correspondence between a range of thresholds and the resulting purity. 

As mentioned above, our choice is to estimate the number of discoveries $V$ from the data itself, namely from the statistics of the local maxima {$\bM^-$}, since their distribution is the same as \dm{that} of $\dm{\bM^+}$  when only noise is present. Let us denote by $N^-$ the number of voxels found in the background region\footnote{For \udft, $N^-\approx 2.4\times 10^8$, which represents $\approx 64\%$ of the total number of voxels $N$ \dm{($N\approx3.8\times 10^8$)}.} and by $V^-(\gamma)$ the number of local maxima of $\bM^-$ with values larger than a given threshold $\gamma>0$ in that region. If $\bM^+$ was obtained from  a noise process with the same statistics as $\bM^-$, an estimate of the number of false discoveries $\widehat{V}$ that would be obtained by
thresholding the full data cube $\bM^+$ (entailing $N$ voxels) 
at a value $\gamma$ is 
\beq
\widehat{V}(\gamma)=\frac{N}{N^-} V^-(\gamma).
\eeq
Hence, if we denote by $R(\gamma)$ the number of local maxima above the threshold 
in $\bM^+$, the purity can be estimated as
\beq
\widehat{P}(\gamma):= 1 -\frac{\widehat{V}(\gamma)}{R(\gamma)} = 1 -\frac{N}{N^-} \cdot \frac{V^-(\gamma)}{R(\gamma)}.
\label{purest}
\eeq

This approach is very similar to the Benjamini \& Hochberg procedure for controlling the FDR. The difference is that
the probabilities of local maxima to be larger than some  value under ${\mathcal{H}}_0$ (the $P$-values) are estimated directly from $\bM^-$ instead of relying on a theoretical
distribution  of the local maxima. Fig. \ref{fig:udf_purity} shows that
this procedure allows for an efficient control of the purity. The value of the threshold $\gamma^*$
\dm{such that} $\widehat{P}(\gamma^*)=P^*$ (with $P^*$ the purity chosen by the user) is selected
and $\bM^+$ is thresholded at this value.

\subsection{Pre-detection of bright emission lines}
\label{sec:pre}
The nuisance removal via the PCA described in Sec.\ref{sec:PCA} is aimed at removing
any source that is not a faint unresolved emission line. The counterpart of the  efficiency of Algorithm 1 is that  powerful emission lines are detected by tests $t_2$ in Algorithm 1, so their contribution
is included in matrix $\bN$, captured by the PCA and thus removed from the data
cube. It is thus necessary to make a predetection of bright emission lines. Such
lines are easily detectable by a high peak emission in the residual data cube $\widetilde{\bR}$. The detection procedure for bright emission lines
mirrors that described in Sec.\ref{sec:purity}, simply replacing $\bM^+$ and $\bM^-$ by  the local maxima of $\widetilde{\bR}$ and $-\widetilde{\bR}$. For \udft, the target purity of this stage is set to $P^*=0.95$. 

\subsection{\dm{Line merging and source extraction} }
\label{sec:lines}

Once the detections are available, we need to group them into sources, where a given source can have multiple lines. This can be a tricky step in regions where bright continuum sources are present because such sources can lead to detections at different spatial positions despite the DCT+PCA (see Fig.~\ref{comparcor}). To solve this problem we use the information from a segmentation map (that can be provided or computed automatically on the continuum image to identify the regions of bright or extended sources, cf sec. \ref{sec:spatial}) and we adopt a specific method for detections that are in these areas.

First, the detections are merged based on a spatial distance criteria (parameter called \texttt{tol\_spat}). Starting from a given detection, the detections within a distance of \texttt{tol\_spat} are merged. Then by looking iteratively at the neighbors of the merged detections, these neighbors are merged in the group if their distance to the seed detection is less than \texttt{tol\_spat} voxels, or if the distance on the wavelength axis is less than a second parameter, \texttt{tol\_spec} (that is, if a line is detected almost at the same wavelength but a different spatial position). This process is repeated for all detections that are not yet merged.

Then we take all the detections that belong to a given region of the segmentation map, and if there is more than one group of lines from the previous step we compute the distances on the wavelength axis between the groups of lines. If the minimum distance in wavelength is less than \texttt{tol\_spec}, the groups are merged.

Finally, for each line we then compute a detection image, obtained by summing the GLR datacube on a window centered on the detection
peak wavelength and a width of 2$\times$FWHM, where FWHM is the width of the spectral template that provides the highest 
correlation peak. We also compute the corresponding spectrum by weighted summation over the spatial extent of the line using the detection image as weight.

\section{Application to MUSE HUDF datacube}
\label{sect:udf}
\ori\ was initially developed for the blind search of \lae\ in the MUSE deep exposure of the Hubble Ultra Deep Field (hereafter HUDF). A preliminary version of the code was successfully used for the blind search exploration of the entire HUDF field of view \citep{Bacon2017}. It resulted in the detection of 692 \lae\ \citep{Inami2017}, including 72 not detected in the HST deep broad band images \citep{Bacon2017, Maseda2018}. 

Here we use the latest version of the \udft\ MUSE datacube (see Fig.~1 of \cite{Bacon2017}), the single 1 arcmin$^2$ datacube that achieves a depth of 30 hours. 
In this field \cite{Inami2017} report the detection of 158 \lae, including 30 not detected in the HST deep broad band images.
Compared to the previous version datacube of the same field used in \cite{Bacon2017}, the data benefits from an improved data reduction pipeline, resulting in less systematics. This version is the one used in the MUSE HUDF data release II (Bacon et al, in prep). 
In this section we focus on the use and performance of the algorithm with an emphasis on the \lae\ search.  

\subsection{Processing}
The released version 1.0 of \ori\ was used on the MUSE \udft\ datacube with the 
 parameters as given in Appendix \ref{annex:parameters}. While most of the parameters can be used with their default value, a few need more attention: the probability of false alarm for the PCA ($P_{FA}^{PCA}$, sec.~\ref{sec:PCA}) and the allowed minimum purity ($P^*$, sec.~\ref{sec:purity}). A correct setting of the first parameter  ensures that the signal coming from bright continuum 
sources and the systematics left by the data reduction pipeline are properly removed. When  $P_{FA}^{PCA}$ is too
low the test threshold is too high and too few spectra are cleaned, leaving nuisances in the residual data cube. When $P_{FA}^{PCA}$ is
too large the test threshold is too low and the signal from some emitters  can be impacted. After some trials, a value of 10\% was used (see an example in Fig.~\ref{fig:thresh}). Note that with such a value, 7\% of the  emitters (the brightest ones) are killed by the PCA process, but they are recovered in the last step of the method (sec.~\ref{sec:pre}). 

The second parameter, the target purity $P^*$, impacts the detection threshold above which the local maxima of the test statistics  are considered as detected lines. A purity value of 80\% was selected as a compromise between the number of false and true detections. The impact and the reliability of this parameter on the detection performance  is discussed in later in section \ref{sec:udf_purity}.

We also spent some time to the \dm{design} of an `optimum' input spectral line dictionary (sec.~\ref{sec:test}). After some trials we found that a set of 3 Gaussian profiles with FWHM of 2, 7 and 12 spectral pixels offers both good detection power (completeness) and affordable computational load. A lower number of profiles degrades the detection power  while a higher number does not increase it but requires more computing power. More sophisticated profiles (e.g., asymmetric line shape mimicking those generally found in \lae) were also considered but not used given their negligible impact on the performances.

\subsection{Results}
\label{sec:results}
The algorithm detects 791 emission lines belonging to 446 different sources. The algorithm assigns to each detected emission line a significance level, computed as the maximum purity at which this line can be detected (hence, the significance level of all detected lines is above $P^*$, which is $80$\% here).

After careful inspection by a set of experts, we confirm 248 \lae\ covering a broad range of redshifts (2.8-6.7) and 133 lower redshifts galaxies, mostly \oii\ emitters but also a few nearby \ha\ emitters{\footnote{\dm{ We also detect a few \ciii\; emitters, as well as \oiii\; emitters but only a few with respect to the \oii\; emitters which are ten times more numerous. Note that most of \ciii\; emitters have low equivalent width and strong continuum.}}}.  
Table \ref{tab:successrate} gives the success rate of redshift assignment for the full sample as a function of the purity estimation. Note that the measured success rate is  smaller than the expected purity. This apparent discrepancy is due to the fundamental difference between a line detection and a redshift assignment. The latter is a very complex process involving matching spectral signature with template, searching for multiple lines, measuring line shape and even \dm{performing} deblending of the source from its environment. Thus at low S/N it is not surprising that some of the real detections do not lead to a confirmed redshift.   

\begin{figure*}
    \includegraphics[width=\textwidth]{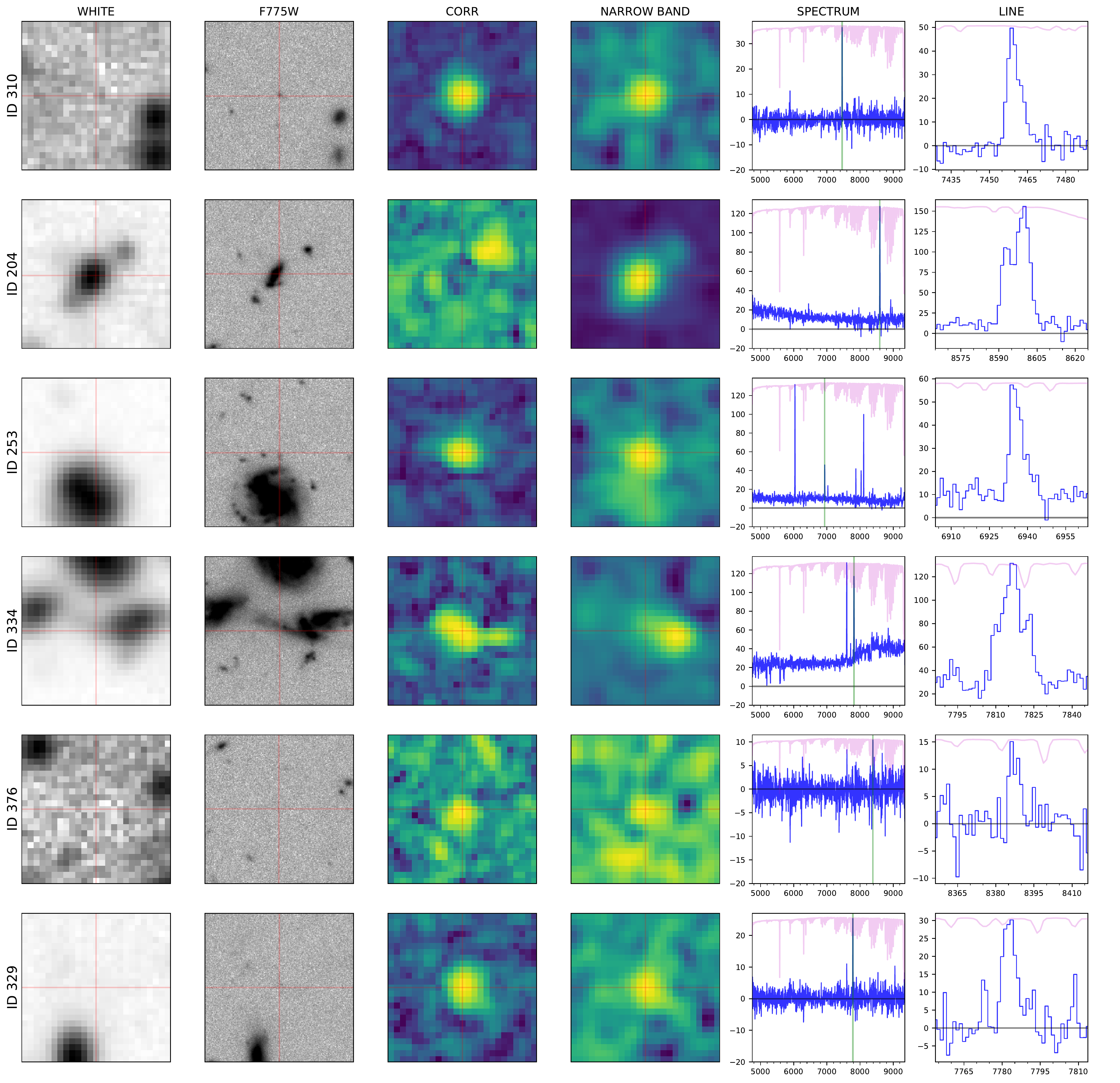}
    \caption{Examples of \ori\ detections in \udft. For each detected source, we show the white light image
    reconstructed from the MUSE datacube (WHITE column), the corresponding HST image in the F775W filter$^\dag$ (F775W column), the detection image obtained by the averaging the GLR datacube over the 
    detected wavelengths (CORR column) and the narrow band image obtained by averaging the raw
    datacube over the same wavelength range (NARROW BAND column). Note that the narrow band image is smoothed with a 0.6\"\ FWHM Gaussian kernel to enhance visually possible sources. The box size of the images is 5 arcsec.\\
    The source spectrum (smoothed with a boxcar of 5 pixels)
    is shown in the SPECTRUM column. The corresponding noise standard deviation is displayed in magenta (inverted on the y axis) and the green line displays
    the wavelength of the \ori\ detection. The last column (LINE) displays the
    (unsmoothed) spectrum zoomed around the \ori\ detected wavelength. 
    }
    $^\dag$\scriptsize The Hubble ACS broad band filter F775W has an effective wavelength of 7624 \AA\ which is nearly aligned with the central wavelength of MUSE (7000\AA). Its effective band pass of 1299\AA\ covers a third of the MUSE band pass (4650-9350\AA).\\
    \label{fig:udf2}
\end{figure*}

A few representative examples of \ori\ detections are given in Fig.~\ref{fig:udf2}. In the following we discuss each case.

The first case (\ori\ ID 310, first row of Fig.~\ref{fig:udf2}) shows the detection of a \lya\ emitter. 
The source cannot be seen in the MUSE white light image but is clearly visible in the HST deep F775W image.
The corresponding \ori\ detection peaked at 7459~\AA. The pseudo narrow band detection image is obtained by averaging the datacube of the GLR test statistics  as described in sec.~\ref{sec:test}, on a window centered on the detection peak (see sec.~\ref{sec:lines}). A similar image
can be obtained by performing the same operation on the raw datacube. As expected the corresponding narrow-band image is more noisy compared to the detection image. However, the smoothed narrow band image displays a clear signal, in line with \ori\ detection. 
The corresponding bright emission line seen in the source spectrum is broad
and asymmetric with a tail on the red side of the peak, a characteristic of \lya\ emission line. The redshift of the source is 5.13.
The HST matched source has the ID 10185 in the Rafelski catalog \citep{Rafelski2016}, its magnitude is 
AB 28.8 in F775W and its Bayesian photometric redshift is 0.61, but with a large error bar (0.47-4.27).
For such a faint source the photometric redshift accuracy is not much reliable, even in this field which has an exquisite suite of deep images spanning a large wavelength range from UV to IR \citep{Brinchmann2017}. This demonstrates the power of a blind detection method like \ori, which does not rely on prior information.

The second case (\ori\ ID 204) shows the detection of a bright \oii\ emitter at z=1.3. The source is bright (F775W AB 24.5) and clearly visible in MUSE white light and HST broad band images. The \oii\ doublet is prominent in the spectrum. While the reconstructed narrow-band image displays a well defined spatial structure that looks like the broad band image, this is not the case of the detection image, which displays a hole at the location of the galaxy center. This is due to the PCA continuum subtraction (sec.~\ref{sec:PCA}), which has removed the brightest \oii\ emission, leaving only the faintest \oii\ lines in the outskirts of the galaxy. In this case the \oii\ emission was recovered in the  predetection stage in the S/N datacube (see sec.~\ref{sec:pre}).

The third source (\ori\ ID 253) is a good example of the power of \ori\ blind detection in the vicinity of a bright continuum object. \ori\ has detected an emission line in the vicinity of the nearby (z=0.62) bright (AB 22) spiral galaxy ID 23794 \citep{Rafelski2016}. The detection and narrow-band images display a coherent structure that spatially matches a small source in the F775W image. Without the \ori\ detection, one would have assumed that this small source in the HST image is one of the H$_2$ region, which can be seen along the spiral arm of the galaxy. The fact that the multi-band HST segmentation map did not identify the source as a different object from the main galaxy would have strengthened this (false) assumption. However, the spectrum confirmed that the \ori\ line does not belong to the main galaxy, but is a \lya\ emitter at z=4.7 with a typical asymmetric line profile. 

The next example (\ori\ ID 334) shows the detection of an emission line in the external part of a galaxy. But contrarily to the previous case, the narrow band and detection images \dm{do} not show the same structure. The narrow band image points to the galaxy core, while the detection image \dm{does} not show much signal there. After inspection of the two spectra, one can demonstrate that the line detected by \ori\ is the H10 Balmer line belonging to the main galaxy spectrum. This line was faint enough to be left by the PCA continuum subtraction. This example shows that care must be taken when analyzing detection results in the region of bright continuum sources. 

The last two examples (\ori\ ID 376 and 329) display detection of faint \lae\ without HST counterpart. 
The sources are clearly visible in the detection images and to some extent also in the narrow band images, but nothing is detected in any HST bands. The spectrum shape is also indicative of \lya\ emission. Given the depth of the HST images (AB $\sim$ 30) in the Hubble UDF, a low redshift object is excluded and the most likely solution corresponds to high redshift \lae. The reality of these detections has been confirmed by \citet{Maseda2018}, who have demonstrated \dm{that} the Lyman-break signal can be recovered by stacking the HST broad-band images of the \ori\ detected \lae\ without HST counterparts.

\begin{table}
\caption{Success rate of redshift assignment for \udft\ \ori\ detections. $Ndetect$ corresponds to the number of sources detected in the considered purity range, $withZ$ (resp. $noZ$) correspond respectively to the number of successful (resp. unsuccessful) redshift assignments.}              
\label{tab:successrate}      
\centering                                     
\begin{tabular}{c c c c c}          
\hline                      
 $Purity$ & $Ndetect$ & $withZ$ & $noZ$ & $Success Rate$ \\    
\hline                                  
    1.00 - 0.95 & 303 & 274  & 29 & 90.4\% \\      
    0.95 - 0.90 & 62  & 28   & 34 & 82.7\% \\
    0.90 - 0.85 & 30  & 11   & 19 & 79.2\% \\
    0.85 - 0.80 & 51  & 13   & 38 & 73.1\% \\
\hline                                             
\end{tabular}
\end{table}

\subsection{Purity and completeness}
\label{sec:udf_purity}

The purity, that is, the fraction of true detections with respect to the total number of detections, is a built-in capability of \ori\ (see sec. \ref{sec:purity}). Obviously we would like to minimize the number of false detections in the total list of detections in order to get a purity as close as possible to 100\%. On the other hand, targeting a higher purity automatically decreases the completeness, that is, the fraction of lines that are detected by the algorithm with respect to the total number of sources to be discovered.
The trade-off between purity and completeness is a feature of any detection method.

The estimation of completeness is highly dependent of the sources we want to detect. For example,  the  completeness is different for a population of unresolved bright \lae ,  or for a population of  unresolved faint \lae\, or for a population of diffuse emission sources with broad emission lines. Hence, a generic estimation of completeness is not a built-in function of \ori\  (in contrast to the estimation of the purity) neither of any detection method.
A detailed study of \lae\ completeness in the HUDF datacubes is beyond the scope of this paper and will be presented in the upcoming DR2 paper (Bacon et al, in prep). Nevertheless, we address here the question of completeness with a simpler approach by generating fake \lae\ into a datacube with similar characteristics to the HUDF datacube. 

In practice we replace the signal of the \udft\ datacube by a random Gaussian noise with zero mean and variance equal to the \udft\ variance estimate. We then generate fake \lae\ using typical number counts representative of the faint end \lya\ luminosity function \citep{Drake2017}. The generated \lya\ lines are asymmetric with  FWHM and skewness that are representative of the \lae\ population. The resulting spectral profiles of the \lae, supposed to be point sources, are convolved with the MUSE PSF. Finally, we add 9 bright continuum sources. The resulting data cube is  an idealized version of the   real data cube but with the same noise characteristics. Note that this process assumes a Gaussian noise distribution, an assumption checked by \cite{Bacon2017} in the \udft\ datacube.

\ori\ is then run on the fake datacube, varying the fluxes, the locations and the wavelengths of the fake \lae. The comparison of the \ori\ detected sources with the input list allows to  compute the true purity  and completeness. 

\begin{figure}
    \centering
    \includegraphics[width=\columnwidth]{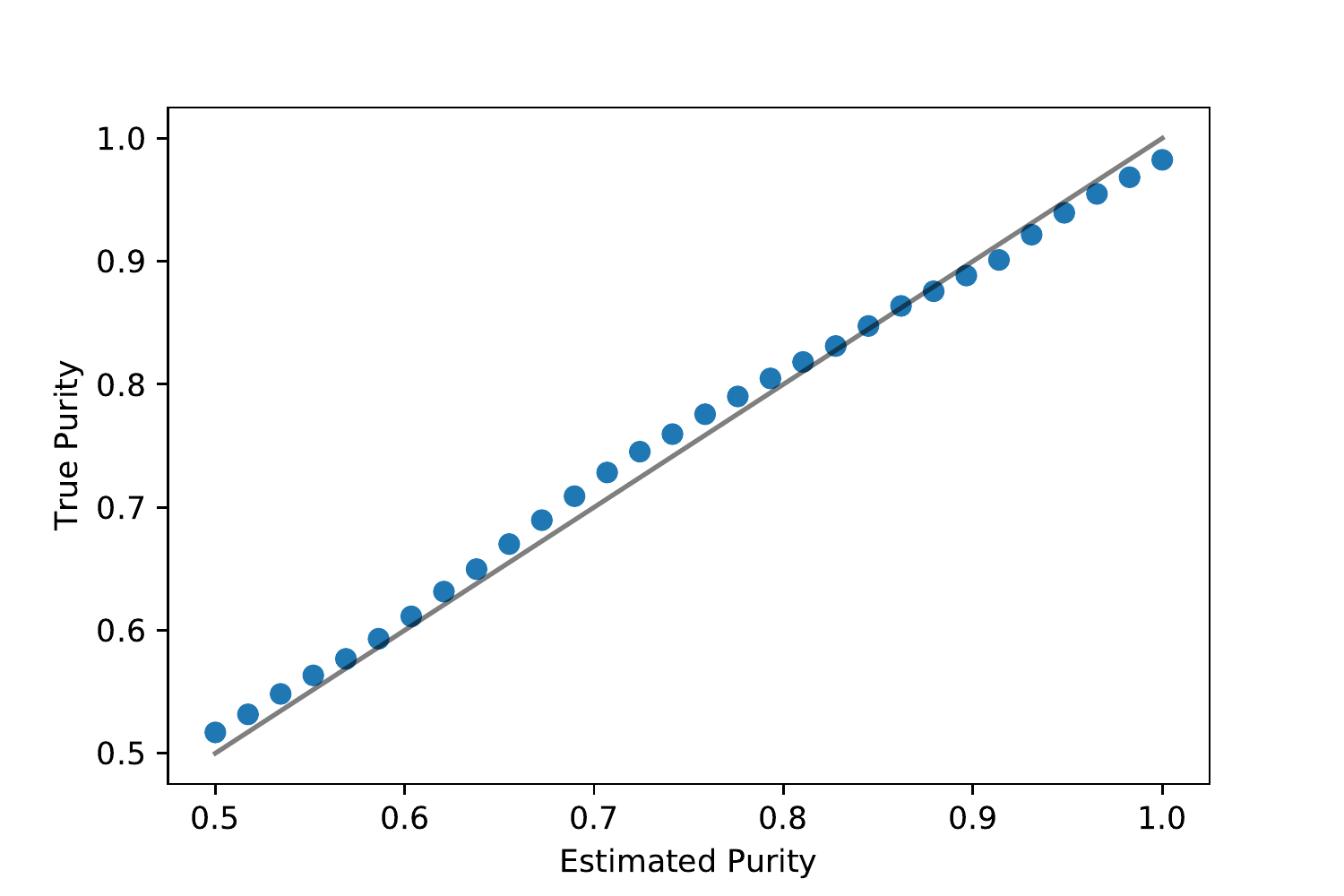}
    \caption{Comparison of the \ori\ estimated purity $\widehat{P}$ (cf Eq.\eqref{purest}) versus the true purity (Eq. \eqref{pur}) for the  \udft\ fake datacube.}
    \label{fig:udf_purity}
\end{figure}

\begin{figure}
    \centering
    \includegraphics[width=\columnwidth]{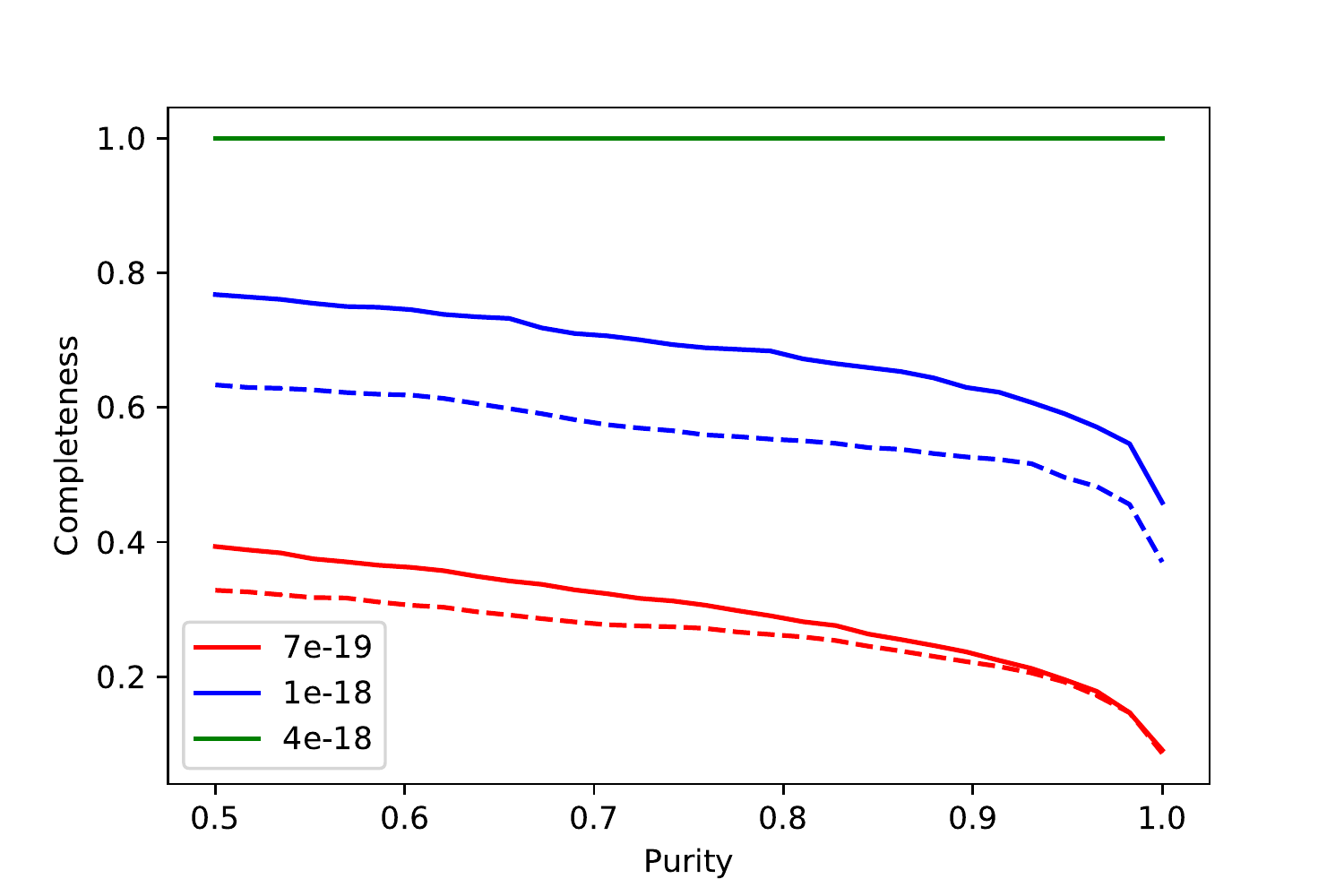}
    \caption{Estimation of completeness versus purity for the \udft\ fake datacube. The values are given for three fluxes (in cgs units) and two wavelength ranges: 6000-7000 \AA\ (solid lines) and 8000-9000 \AA\ (dashed lines).}
    \label{fig:udf_completeness}
\end{figure}

Fig.~\ref{fig:udf_purity} compares the purity estimated by the algorithm with respect to the true purity. This shows that the estimate provided by \ori\ is reliable on a wide range of purity levels.
Fig.~\ref{fig:udf_completeness} shows completeness versus purity plots for two different wavelength ranges and three flux values. As expected bright sources are fully recovered while the algorithm achieves a lower completeness for fainter ones. Note also that completeness is lower in the red for a given flux because of the impact of sky lines (the larger variance in the red end is visible in the magenta plots of Fig.~\ref{fig:udf2}, column SPECTRUM). Finally, note that the  weak slopes of the completeness versus purity curves indicate
that \ori\ is able to achieve a  relatively high completeness (with respect to the `asymptotic' one) with a fairly high purity (0.8 for instance).

\subsection{Discussion}

With respect to the preliminary, unreleased version of \ori\ used on the version 0.42 of the \udft\ datacube in \cite{Bacon2017}, we have increased the number of detections of \lae\ by 34\% (255 versus 190). Note that the difference is only 7\% (174 versus 163) when we restrict the sample to the highest confidence redshifts. This was expected as the improved performance of \ori\ has allowed the detection of fainter \lae\, while the `easiest one' were already identified in the preliminary version of the algorithm.

We also report the detection of 86 \lae\ without HST counterpart. This is almost 3 times more sources with respect to the 30 sources published in \cite{Bacon2017}. The ability to detect such sources, which are invisible in the deepest HST broad band images, is an important feature of \ori. 

While some of these improvements are due to the better data quality of the latest version of the datacube, most of them result from the advanced methodological features of the present version of \ori\ with respect to the preliminary version used in \cite{Bacon2017}. The most important improvements regard the continuum subtraction and the use of robust test statistics. 
Last, but not least, \ori\ gives now an important reliability factor, the estimated purity, a must for a robust detection method. 

Although the DCT and iterative PCA do a good job for the continuum subtraction (in particular better than a basic median filtering, see sec.~\ref{sec:PCA}), it still leaves some low level residuals that can produce spurious detections (such an example -- \ori\ ID $334$-- is highlighted in sec.\ref{sec:results}). The consequence is that the purity in the regions corresponding to bright continuum sources is lower than in the background region. A precise estimate of the purity in that case is difficult.

Among the many results brought by the analysis of the MUSE deep field observations is the ubiquitous presence of \lya\ low surface brightness extended halos surrounding \lae\ \citep{ Wisotzki2016, Leclercq2017, Wisotzki2018}. Given that \ori\ was designed for point source detection one can ask whether  a better strategy would have been to use a  kernel larger than the PSF. Even if the total flux of the \lya\ halo can be similar to or even larger than the flux of the central component, its surface brightness is approximately two magnitude lower than the central component outside the PSF area (see Fig.~2 of \citealt{Leclercq2017}). Hence, in practice, the point like assumption does not appear to be an important limiting factor for the blind detection of \lae .

\section{Summary and conclusion}
 The
algorithm \ori\ is designed to be as powerful as possible for detecting faint spatial-spectral emission signatures, to allow for a stable false detection rate over the data cube and to provide an automated and reliable estimation of the resulting purity.
We have shown on  realistic simulated MUSE data  that the algorithm achieves these goals and \ori\ was applied to the deepest MUSE observations of the Hubble Ultra Deep Field (\udft). In this tiny 1\arcmin $\times$ 1\arcmin\ field, \ori\ revealed a large population of \lae\ (255),  including 86 sources without HST counterpart. These sources  could not have been found without such a powerful blind detection method as \ori.

While the algorithm is mostly automated, we have presented a list of the main  parameters with guidelines allowing to tune their default values. The algorithm is freely available to the community as a Python package on GitHub.

We have already identified points for improvements to the current version of \ori\ and we are currently working on them. The first point is the greedy PCA (Sec. \ref{sec:PCA}). Although this step works quite well and is fast, it may still leave residual nuisances that increase the false detection rate in the region of bright extended sources. Besides, we have not analyzed the behavior of this step in
different acquisition settings, for instance crowded fields or fields with much lower signal-to-noise ratios than \udft. Finally, we are  developing a method for automatically tuning of the $P_{FA}^{PCA}$ parameter involved in this stage.

We also know that there is  room for improvement in the detection power (and completeness) of the method, by accounting for a more complex model than model \eqref{mod} (leading then to GLR test statistics different from   \eqref{tglr}). We have already devised 
slightly more powerful models and tests' statistics, but those   make the total amount of computing power far too demanding, as they  impose
a processing subcube per subcube (instead of allowing for fast convolutions along the spatial and spectral dimensions). Improvements might yet be found in this direction in the future.

Finally, the estimation of the purity can also be improved. While the current estimation is
efficient for  data cubes with extended zones of `background', we know that the procedure may
fail for crowded data cubes where no or too few such zones exist. In such 
 cases, the amount of  data available to estimate the test statistics under the null hypothesis is insufficient,
 a problem which is the object of active research in statistics.

\begin{acknowledgements}
 DM made extensive use of the free software GNU Octave \citep{octave} for developping \ori\ and is in particular thankful to the developers of the Octave packages \textit{statistics}, \textit{signal} and \textit{image}. DM also acknowledges support from the GDR ISIS through the \textit{Projets exploratoires} program (project TASTY).  Part of this work was granted access to the HPC and visualization resources of the Centre de Calcul Interactif hosted by University of Nice C\^ote d'Azur.
RB acknowledges support from the ERC advanced grant 339659-MUSICOS. 
This research made use of the following open-source packages for Python and we are thankful to the developers of these: Astropy \citep{astropy:2013, astropy:2018}, Matplotlib \citep{matplotlib}, MPDAF \citep{Piqueras2016}, Numpy \citep{numpy}, Photutils \citep{photutils}, Scipy \citep{scipy}.
\end{acknowledgements}

\bibliographystyle{aa}
\bibliography{biblio.bib}

\begin{appendix}
\section{Mathematical description}

\subsection{Derivation of $T_{GLR}^+$}
\label{deriv}
 The Generalized Likelihood Ratio computes
the ratio of the likelihoods under both hypotheses, with the unknown parameters 
set as their maximum likelihood estimates. For the composite model \eqref{mod} this
leads to
\beq
\textrm{GLR:}\quad \frac{\displaystyle{\max_{i,a,b}} p({{\bf{f}}; i, a, b)}}{\displaystyle{\max_{a}} p({({\bf{f}};  a)}}.
\label{glr1}
\eeq
For $i$ fixed, this problem has been considered by \cite{scharf94}. From their expression
(5.15), where ${\bf{x}}, {\bf{S}}$ and ${\bf{y}}$ correspond respectively to
${\bf{d}}_i, {\bf{1}}$ and ${\bf{f}}$, the GLR test statistic is
\beq
T_i:=\max\left( 0, \frac{{\bf{f}}^\top \textrm{P}_{{\bf{1}}}^\perp {\bf{d}}_i}  {({\bf{d}}_i ^\top \textrm{P}_{{\bf{1}}}^\perp {{\bf{d}}_i})^{1/2}} \right),
\label{lego}
\eeq
where $\textrm{P}_{{\bf{1}}}^\perp$ denotes the projection on the orthogonal complement of ${{\bf{1}}}$:
\beq
\textrm{P}_{{\bf{1}}}^\perp:= {\bf{I}} - {{\bf{1}}}({{\bf{1}}}^\top {{\bf{1}}})^{-1} {{\bf{1}}}^\top = {\bf{I}}-\frac{{{\bf{1}}}{{\bf{1}}}^\top}{n_f},
\eeq
with $n_f=n_x n_y n_z$ (the number of voxels in {\bf{f}}).
 Now note that
 \beq
 \begin{cases}
 \textrm{P}_{{\bf{1}}}^\perp {\bf{d}}_i = {\bf{d}}_i -  \frac{{{\bf{1}}}{{\bf{1}}}^\top}{n_f}  {\bf{d}}_i
 = {\bf{d}}_i -\frac{1}{n_f}( {\bf{d}}_i^\top {{\bf{1}}}) {{\bf{1}}} := {\overline{\bf{d}}}_i,\\
 {\bf{f}}^\top \textrm{P}_{{\bf{1}}}^\perp {\bf{d}}_i= {\bf{f}}^\top {\overline{\bf{d}}}_i.
 \end{cases}
 \eeq
 Hence, the GLR test statistic for fixed $i$ \eqref{lego} can be rewritten as
 \beq
T_i=
\max \left( 0,
\frac{{\bf{f}}^{\;\top} \overline{\bd} _i}{\|  \overline{\bd}_i \|} \right),
\label{lego2}
\eeq
and computing the maximum over the index $i$ as in \eqref{glr1} leads to
\beq
T_{GLR}^+ = \max_i T_i = \max_i\left( 0,
\frac{{\bf{f}}^{\;\top} \overline{\bd} _i}{\|  \overline{\bd}_i \|} \right).
\eeq
\subsection{Statistics of $T_{GLR}^-$}
\label{app1}
The statistics of $T_{GLR}^-$ can be derived using an approach similar to the Proposition II.1 of \cite{DBLP:journals/tsp/BacherMCM17}, which we report and adapt here as our setting is slightly different in the considered model \eqref{mod} and test statistics \eqref{tglr}. First, as \cite{DBLP:journals/tsp/BacherMCM17} we  make the hypothesis that under ${\mathcal{H}_0}$ the noise is symmetric. We do not require the noise to be centred as the considered GLR tests statistics is invariant to an  arbitrary shift. Second, note that for any subcube ${\bf{f}}$,
\beq
\frac{{\bf{f}}^{\;\top} \overline{\bd} _i}{\|  \overline{\bd}_i \|}
= -\frac{(-{\bf{f}})^{\;\top} \overline{\bd} _i}{\|  \overline{\bd}_i \|}.
\label{sym}
\eeq
In words, the value of the amplitude estimated  when fitting a line profile to a spectrum is the opposite of the value obtained when fitting the profile to the opposite spectrum.
Third, note that for any finite set of real numbers $\{a_i\}$,
\beq
\max_i \left(0,a_i\right)=-\min_i \left(0,-a_i\right).
\label{maxj}
\eeq
Consider now $T_{GLR}^-$ in \eqref{glrm}.  We have
\beqna
\max_i\left(0,\frac{{\bf{f}}^{\;\top} \overline{\bd} _i}{\|  \overline{\bd}_i \|} \right)& = &\max_i\left(0,-\frac{-{\bf{f}}^{\;\top} \overline{\bd} _i}{\|  \overline{\bd}_i \|} \right)\\
&=& -\min_i\left(0,\frac{-{\bf{f}}^{\;\top} \overline{\bd} _i}{\|  \overline{\bd}_i \|} \right)
\eeqna
where the first equality comes from \eqref{sym} and the second from \eqref{maxj}. 
Since under ${\mathcal{H}}_0$, ${\bf{f}}$ and $-\bf{f}$ have the same distribution,
the second equality above shows that $T_{GLR}^+$ and $T_{GLR}^-$
 also share the same distribution.

\section{Implementation}
\label{sect:implementation}

ORIGIN was developped in GNU Octave   with a twin version ported and optimized in Python as the Python package \texttt{muse\_origin}. Its source code is available on GitHub\footnote{https://github.com/musevlt/origin} under a MIT License. A complete documentation is also available on readthedocs\footnote{https://muse-origin.readthedocs.io/en/latest/}.

As the processing of a MUSE data cube with the ORIGIN algorithm is relatively complex, it is divided in steps corresponding roughly to the steps described in Section~\ref{sect:stepbystep}.  Each step produces intermediate results. This allows to stop at a given point and to inspect the results. It is possible to save the outputs after each step and to reload a session to continue the processing.

To run ORIGIN, it is first necessary to instantiate a \texttt{muse\_origin.ORIGIN} object. In the considered
framework this object is a MUSE data cube, which usually contains information about the FSF (if not, this information must be provided separately). The  name given to this object  is the \texttt{session} name used as the directory name in which outputs are  saved. (By default this is inside the current directory but this can be overridden with the \texttt{path} argument)).

Here is an example:

\begin{verbatim}
>>> from muse_origin import ORIGIN
>>> orig = ORIGIN(CUBE, name='origtest')
INFO : Step 00 - Initialization (ORIGIN v3.2)
INFO : Read the Data Cube minicube.fits
INFO : Compute FSFs from the datacube
INFO : mean FWHM of the FSFs = 3.32 pixels
INFO : 00 Done
\end{verbatim}

The processing steps described in this article  can the be run on this object. For instance for the first step this yields:

\begin{verbatim}
>>> orig.set_loglevel('INFO')
>>> orig.step01_preprocessing()
INFO : Step 01 - Preprocessing
INFO : DCT computation
INFO : Data standardizing
INFO : Std signal saved in self.cube_std ...
INFO : Compute local maximum of std cube values
INFO : Save self.cube_local_max ...
INFO : DCT continuum saved in self.cont_dct ...
INFO : Segmentation based on the continuum
INFO : Found 11 regions, threshold=1.94
INFO : Segmentation based on the residual
INFO : Found 3 regions, threshold=1.12
INFO : Merging both maps
INFO : Segmap saved in self.segmap_merged ...
INFO : 01 Done - 2.50 sec.
\end{verbatim}

The detailed contents of each step are described in more details in the documentation\footnote{https://muse-origin.readthedocs.io/}, which also contains an  Jupyter notebook example. 

The other steps can be run with the following commands:

\begin{verbatim}
>>> orig.step02_areas()
>>> orig.step03_compute_PCA_threshold()
>>> orig.step04_compute_greedy_PCA()
>>> orig.step05_compute_TGLR()
>>> orig.step06_compute_purity_threshold()
>>> orig.step07_detection()
>>> orig.step08_compute_spectra()
>>> orig.step09_clean_results()
>>> orig.step10_create_masks()
>>> orig.step11_save_sources()
\end{verbatim}

Each step produces various files, which are useful to analyze the algorithm's behavior and the influence of its parameters. The most important files are the final catalogs with the list of all the detected lines (\texttt{orig.Cat3\_lines}) and sources (\texttt{orig.Cat3\_sources}). The last step also creates MPDAF \emph{Source files}\footnote{https://mpdaf.readthedocs.io/en/latest/source.html}, which gather all the information for each source (GLR cube, images, masks, spectra, cf sec. \ref{sec:lines}).

We underline that a substantial amount of time was devoted to optimize as much as possible the numerical implementation of the considered methods in order to minimize the overall computation time. Table \ref{tab:computationtime} gives the computation times for each step for the \texttt{udf10} data cube described in Section \ref{sect:udf}. These times were obtained on a computing machine with 80 Intel\textregistered\ Xeon\textregistered\ Gold 6148 CPU @ 2.40GHz CPUs, and using the optimized version of Numpy with the Intel\textregistered\ MKL. As ORIGIN uses intensively linear algebra for the iterative PCA and FFT for the profiles convolution, using the Intel\textregistered\ MKL with Numpy may bring significant performance boost. This is the case by default when using Anaconda or Miniconda, and it is also possible to use the Numpy package from Intel\textregistered\ with \texttt{pip}. Intel\textregistered\ also provides an \texttt{mkl-fft}\footnote{https://github.com/IntelPython/mkl\_fft} package with a parallelized version of the FFT.


\begin{table}
\caption{Computation times for each step (minutes) for the \texttt{udf10} data cube described in Section \ref{sect:udf}.}
\label{tab:computationtime}      
\centering                                     
\begin{tabular}{lc}
\hline
Step & Execution Time \\
\hline
\texttt{step01\_preprocessing} & 01:49 \\
\texttt{step02\_areas} & 00:02 \\
\texttt{step03\_compute\_PCA\_threshold} & 00:03 \\
\texttt{step04\_compute\_greedy\_PCA} & 02:47 \\
\texttt{step05\_compute\_TGLR} & 05:31 \\
\texttt{step06\_compute\_purity\_threshold} & 00:12 \\
\texttt{step07\_detection} & 00:05 \\
\texttt{step08\_compute\_spectra} & 02:01 \\
\texttt{step09\_clean\_results} & 00:16 \\
\texttt{step10\_create\_masks} & 00:13 \\
\texttt{step11\_save\_sources} & 01:45 \\
\hline
Total & 14:49 \\
\hline
\end{tabular}
\end{table}

\newpage
\newpage

\section{\ori\ parameters}
\label{annex:parameters}
\label{table}
\newpage
In the following table we give the \ori\ parameters used for \udft\ processing:
\bigskip

    \begin{tabular}{c|c|c|c|l}
    \hline
      Step in Sec. & Symbol & Python variable & Value & Description \\
      \hline
      \ref{sec:dct} &   $N_{DCT}$ & dct\_order & 10 & Order of the DCT\\
      \hline
      \ref{sec:spatial} &   $P_{FA}^{seg}$ & pfa & 0.2 & ``False alarm'' rate for tests \eqref{test_t1t2}\\
      \hline
       \ref{sec:spatial} &   $S_{min},S_{max}$ & minsize,maxsize & 80,120 & Min and max surface for the large segmentation patches \\
      \hline
       \ref{sec:PCA} &  $ P_{FA}^{PCA}$ & pfa\_test & 0.1 & ``False alarm'' rate for test $t_2$ in \eqref{test_t1t2} during PCA cleaning \\
        \hline
       \ref{sec:PCA} &  $ F_b$ & Noise\_population & 0.05  & Fraction of spectra of $\bB$ used to estimate background \\
        \hline
       \ref{sec:purity} &  $N_s$ & & 3 & Number of spectral profiles for the detection \\
         \hline
       \ref{sec:test} &  $P^*$ & purity & 0.8 & Target purity for faint emission lines \\
         \hline
       \ref{sec:pre} &  $P^*$ & purity\_std & 0.95 & Target purity for bright emission lines \\
      \hline
       \label{udf:tab1}
    \end{tabular}

\end{appendix}

\end{document}